\documentclass[twocolumn,nofootinbib,preprintnumbers,longbibliography,amsmath,amssymb,aps]{revtex4-2}
\usepackage{makecell}
\usepackage[pdftex]{graphicx}
\usepackage{dcolumn}
\usepackage{bm}
\usepackage{float}
\usepackage{hyperref}
\usepackage{epstopdf}
\usepackage{mathrsfs}
\usepackage{amsmath}
\usepackage{graphicx}
\usepackage{multirow}
\usepackage{xr-hyper}
\newcommand{\ket}[1]{\left| #1 \right>}

\newcommand{\Fig}[1]{Fig.\,\ref{#1}}
\newcommand{\Eq}[1]{Eq.\,\eqref{#1}}

\hypersetup{
            colorlinks=true,
            linkcolor=blue,
            anchorcolor=blue,
            citecolor=blue}
\begin{document}

\title{Magnon spin photogalvanic effect induced by Aharonov-Casher phase}  

\author{YuanDong Wang$^{1,2}$}

\author{Zhen-Gang Zhu$^{1,2,3}$}
\email{zgzhu@ucas.ac.cn}

\author{Gang Su$^{2,3,4}$}
\email{gsu@ucas.ac.cn}

\affiliation{$^{1}$ School of Electronic, Electrical and Communication Engineering, University of Chinese Academy of Sciences, Beijing 100049, China.\\
$^{2}$ School of Physical Sciences, University of Chinese Academy of Sciences, Beijing 100049, China. \\
$^{3}$ CAS Center for Excellence in Topological Quantum Computation, University of Chinese Academy of Sciences, Beijing 100049, China.\\
$^{4}$ Kavli Institute for Theoretical Sciences, University of Chinese Academy of Sciences, Beijing 100190, China.
 }

\begin{abstract}
Magnons are electrically neutral bosonic quasiparticles emerging as collective spin excitations of magnetically ordered materials, and play a central role in
the next-generation spintronics owing to its obviating Joule heating. 
A difficult obstacle for quantum magnonics is that  
%
the magnons do not couple to the external electric field directly so that a direct electric manipulation via  bias or gate voltage as in conventional charge-based devices seems not applicable. 
%
%
%
In this work, we propose a new mechanism in which magnons can be excited and controlled by electric field of light directly. Since the electric field of light can be tuned in a wide and easy way, the proposal is of great interest in realistic applications. 
We call it as the magnon spin photogalvanic effect (SPGE), which comes from five contributions: the Drude, Berry curvature dipole (BCD), injection, shift, and rectification, with distinct geometric origins.  We further show that the responses to linearly-polarized or circularly-polarized light are determined by band-resolved  quantum metric or Berry curvature, the two combined together just comprise of a quantum geometric tensor. The proposed magnon SPGE can be measured by a characterized topological phase transition. We also discuss a breathing kagome-lattice model of ferromagnets and suggest possible candidate materials to implement it.
\end{abstract}
\maketitle


\section{Introduction}

Spin transport plays a central role in the field of spintronics \cite{RevModPhys.76.323, bader2010spintronics}. Currently there are significant interests to explore the magnons in magnetic materials.
The magnons, as collective spin excitations of magnetically ordered materials,
are electrically neutral bosonic quasiparticles. The transport of magnons obviates Joule heating at a fundamental level, and thus magnons are regarded as the candidate in pursuing the next-generation spintronics \cite{chumak2015magnon, RevModPhys.90.015005}. For this purpose, a deeper understanding of the properties of magnons and the precise manipulation of magnons are urgently called for. 
However, due to the electrical neutrality of magnons, their coupling to the external electric field is less considered. Therefore, much progress has been made in studying the thermal control on magnons \cite{PhysRevLett.115.266601, PhysRevLett.116.097204, PhysRevLett.104.066403, onose2010observation, PhysRevLett.106.197202, PhysRevLett.117.217202, PhysRevLett.117.217203,  PhysRevResearch.2.013079}.  
However, the thermal control is not easily accurately controlled and sometimes even cumbersome.

One of the most promising approaches is to use optical method to control  magnons. Since magnons possess spin degree of freedom, it has been proposed that the dc magnon spin photocurrent (MSPC) can be generated in antiferromagnets by the Zeeman coupling of magnon spin and the magnetic field component of light, where the angular momentum transfer between  magnons and photons is invoked by applying a circularly polarized (CP) light \cite{PhysRevB.98.134422}. Using the magnetic field component,  the generation of the MSPC via linearly polarized (LP) light has been proposed recently \cite{PhysRevB.100.224411, PhysRevLett.129.107201}. 
 A desired goal is to achieve an enhanced MSPC by the electric field component of the light. For example, the MSPC driven by CP light was proposed via a two magnon Raman process with the coupling to the electric field \cite{PhysRevB.104.L100404}. It is a direct spin angular momentum transfer process that the right-handed photons turn into the left-handed photons by imparting spin angular momentum and creating a magnon pair carrying a net spin current. Therefore, the current is proportional to the chirality of the incident light and vanishes if the light is linearly polarized. The magnetoelectric coupling generally exists in multiferroic materials, for which the magnon can be generated and controlled by electric field \cite{takahashi2012magnetoelectric,PhysRevLett.106.057403,bordacs2012chirality}.  Recently, the MSPC has been predicted in collinear antiferromagnets via the coupling between electric field and polarization with a broken inversion symmetry \cite{PhysRevB.107.064403}. Nevertheless, the spin photocurrent induced by the  interaction between the electric field and the magnons has not been fully understood, and it is unclear that how it is affected by the quantum geometry of magnon bands \cite{mcclarty2022topological}.

In this paper, we develop a general formalism for the MSPC mediated via the Aharonov-Casher (AC) effect of magnons induced by electric field. 
It can be understood that magnons moving in an electric field acquire a geometric phase through the AC effect during the processing as \cite{PhysRevLett.53.319,  PhysRevA.65.013607, PhysRevLett.106.247203} 
\begin{equation}\label{ac-phase}
\theta_{ij}=\frac{g\mu_{B}}{\hbar c^2}\int_{\bm{r}_i}^{\bm{r}_j}(\bm{E}(t)\times \hat{\bm{e}}_{z})\cdot d\bm{r},
\end{equation}
where  $\bm{E}(t)$ is the electric field of light. 
In Eq.(1) we assume the magnetization is along $z$-direction and then the magnetic moment of magnon points $-z$-direction with $\bm{\mu}=-g\mu_B \hat{\bm{e}}_z$, where $g$ is the Land\'{e} factor and $\mu_B$ is the Bohr magneton.
 This phase ($\theta_{ij}$) will enter into the coupling between spins or the hopping integral of electron operators (see Appendix. \ref{appa} and \ref{appb}). The Hamiltonian is thus modified in a form that the effect of electric field of light will play as a driving force. Since this electric field couples to the local magnetization, magnons (variations of magnetization) will be excited consequently.

\section{Theoretical framework}

\subsection{Light-magnon coupling and density matrix equations of motion}
For a general two-body spin interaction Hamiltonian in the absence of external field,
\begin{equation}\label{ss-hami}
H_0=\frac{1}{2}\sum_{i,j}^{L}\sum_{n,m}^{N}\sum_{\alpha\beta}{S}^{\alpha}_{i,n}H^{\alpha\beta}_{nm}(i-j){S}^{\beta}_{j,m},
\end{equation}
where $\bm{S}_{i,n}$ is a spin operator at the $n$th sublattice (with total number $N$) of the $i$th magnetic unit cell (with total number $L$), with $H^{\alpha\beta}_{nm}(i-j)$ the magnetic exchange interaction. The classical ground state is identified by treating the quantum mechanical spin operators as classical vectors and minimizing the classical ground-state energy.
 By setting a global (reference) coordinates $(\hat{\bm{x}},\hat{\bm{y}},\hat{\bm{z}})$, the local coordinates (spherical coordinates) of each spin relate the global coordinate through
$
\bm{S}_{i,n} = R_{n}(\theta_i,\phi_i)\bm{S}_{0}.
$
 The magnons are the usual low-energy excitations in ordered magnets, which is considered via the Holstein-Primakoff transformation in local coordinates \cite{ toth2015linear}
\begin{equation}
\begin{aligned}
&S_{i,n}^{\theta}=\sqrt{\frac{S}{2}}(a_{i,n} + a_{i,n}^\dagger), \quad S_{i,n}^{\phi} = -i\sqrt{\frac{S}{2}}(a_{i,n}-  a_{i,n}^{\dagger}),\\
& S_{i,n}^{r} = S - a_{i,n}^\dagger a_{i,n}.
\end{aligned}
\end{equation}
and we obtain
\begin{equation}\label{sin}
\bm{S}_{i,n}^{\alpha}=\sqrt{\frac{S}{2}}\hat{\bm{u}}_{n} a_{i,n}+\sqrt{\frac{S}{2}}\hat{\bm{u}}_{n}^{ *}a_{i,n}^\dagger + \hat{\bm{z}}_{n}(S - a_{i,n}^\dagger a_{i,n}),
\end{equation}
where $\alpha = x,y,z$, the coefficients $\hat{\bm{u}}_{n}$ and $\hat{\bm{z}}_{n}$ are related to the relative rotation between the global and local coordinates. Using \Eq{sin} and transform into the momentum space, the magnon Hamiltonian is written as $H=\frac{1}{2}\sum_{\bm{k}}\Psi_{\bm{k}}^{\dagger}\mathcal{H}(\bm{k})\Psi_{\bm{k}}$, with the kernel $\mathcal{H}(\bm{k})$ 
 being a 2N$\times$2N bosonic Bogoliubov-de Gennes (BdG) Hamiltonian with the vector boson operator
where $\Psi_{\bm{k}}^{\dagger} = (a_{{\bm{k}},1}^{\dagger}, \cdots , a_{{\bm{k}},N}^{\dagger}, a_{{-\bm{k}},1}, \cdots,  a_{{-\bm{k}},N})$ (for the details of $\mathcal{H}(\bm{k})$ see Appendix. \ref{appa}).

In general the bosonic Hamiltonian $\mathcal{H}(\bm{k})$ does not conserve the particle number, for example, the ferromagnets with elliptical magnons (where an anisotropy deforms the formerly circular precession) or in non-ferromagnets. And the Hamiltonian is diagonalized with the Bogoliubov transformation
\begin{equation}
U^{\dagger}_{\bm{k}}\mathcal{H}_{\bm{k}}U_{\bm{k}}=\mathcal{E}_{\bm{k}},
\end{equation}
which satisfies $U_{\bm{k}}\Sigma_z U_{\bm{k}}^{\dagger} = \Sigma_z $, where the diagonal matrix $\Sigma_z = \text{diag}(1,1,\cdots, -1, -1, \cdots)$ with $N$ positive ones and $N$ minus ones along the diagonal.  The vector boson operator transforms as  $\Psi_{\bm{k}}=U_{\bm{k}}\Phi_{\bm{k}}$. The $m$th column vector encoded in the matrix $U_{\bm{k}}$ stands for the (periodic part of) Bloch wave function for the $m$th magnon band \cite{PhysRevB.87.174427}.  Noting that $\Phi_{\bm{k}}$ does not satisfy the commutation relation of Bosons. Instead it satisfies
\begin{equation}\label{commu}
[\Phi_{\bm{k}},\Phi_{\bm{k}^\prime}^\dagger]=\Sigma_z \delta_{\bm{k},\bm{k}^\prime}.
\end{equation}
The equilibrium density matrix in band space is given as
\begin{equation}
\rho_{\bm{k}m }^{(0)}(t)\equiv \langle (\Phi_{\bm{k}}^{\dagger})_{m}(t)(\Phi_{\bm{k}})_{m}(t) \rangle_{0},
\end{equation}
here the subscript ``0" denotes the equilibrium state, and $(\Phi_{\bm{k}}^{\dagger})_{m}$ is the $m$th element of the vector $\Phi_{\bm{k}}^\dagger$. For later convenience, we write $(\Phi_{\bm{k}}^{\dagger})_{m}$ as $\Phi_{\bm{k}m}^{\dagger}$.
 By using of \Eq{commu}, one obtains
\begin{equation}
\rho_{\bm{k}m} = \begin{cases} g(\mathcal{E}_{\bm{k}m}), & [\Phi_{\bm{k}m},\Phi_{\bm{k}m}^\dagger] = 1 \\ -g(-\mathcal{E}_{\bm{k}m}), & [\Phi_{\bm{k}m},\Phi_{\bm{k}m}^\dagger] = -1,\end{cases}
\end{equation}
where $g(\mathcal{E}_{\bm{k}m})$ is the Bose-Einstein distribution $g(\mathcal{E}_{\bm{k}m})= 1/(e^{\beta \mathcal{E}_{\bm{k}m}} - 1)$, which is short noted as $g_m$ in the following. 
It is convenient to introduce the matrix $\varepsilon_{\bm{k}}$ \cite{fujiwara2022nonlinear}:
\begin{equation}
\varepsilon_{\bm{k}}=\Sigma_z \mathcal{E}_{\bm{k}} = U_{\bm{k}}^{-1}\Sigma_z \mathcal{H}_{\bm{k}}U_{\bm{k}},
\end{equation}
and the density matrix is simplified as $\rho_{\bm{k}m} = \Sigma_{z,mm}g(\varepsilon_{\bm{k}m})$. For general operator, with the Bogolyubov's representation, it transforms as
 \begin{equation}\label{trans}
\hat{\mathcal{O}} = \sum_{\bm{k}} \Psi_{\bm{k}}^\dagger \mathcal{O}_{\bm{k}} \Psi_{\bm{k}} = \sum_{\bm{k}} \Phi_{\bm{k}}^\dagger \Sigma_z \tilde{\mathcal{O}}_{\bm{k}} \Phi_{\bm{k}},
 \end{equation}
with the definition $\tilde{\mathcal{O}}_{\bm{k}}=U_{\bm{k}}^{-1}\Sigma_z \mathcal{O}U_{\bm{k}}$.

\begin{figure}[bt]
\centering
\includegraphics [width=1\columnwidth]{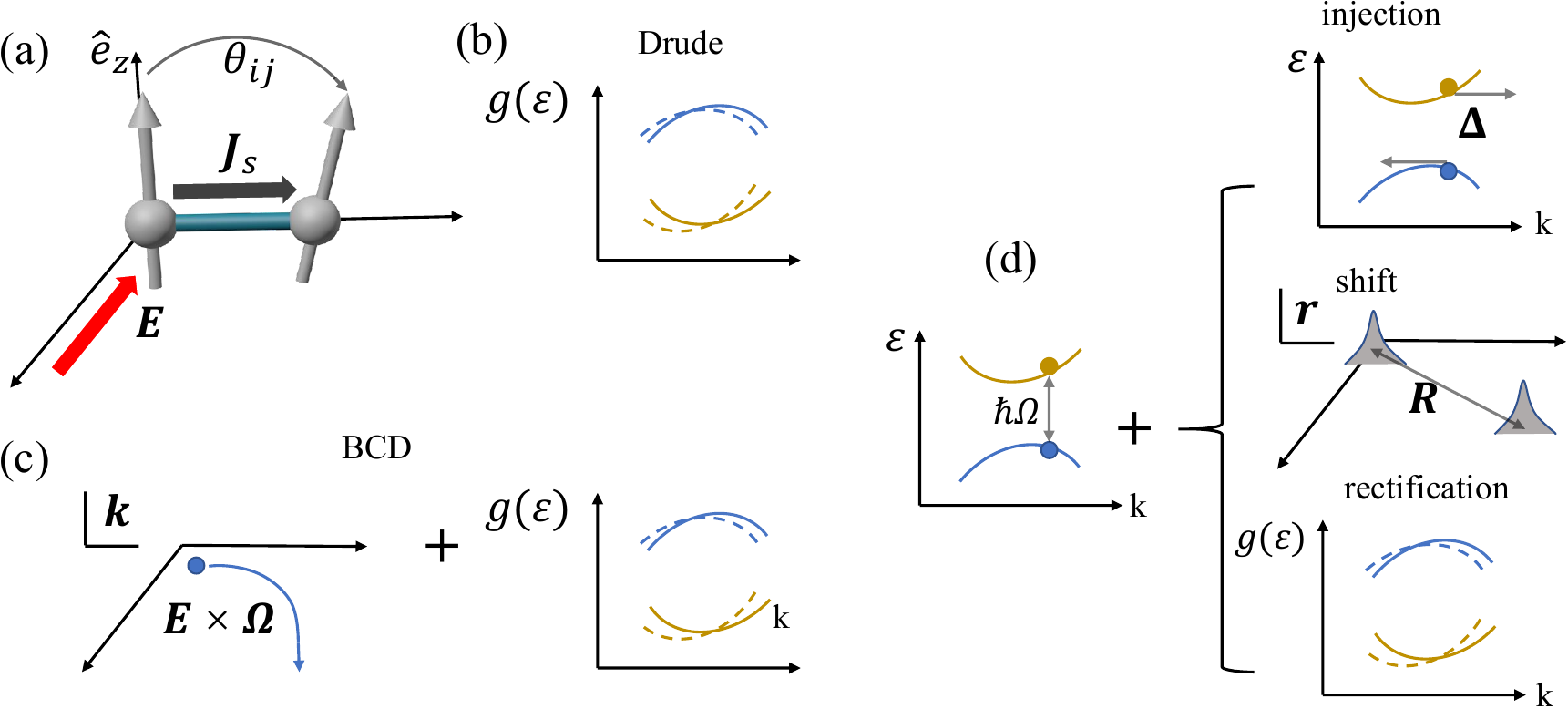}
\caption{\fontsize{5.5bp}{1bp}(a) A schematic of the magnon spin current generated via AC effect with time-dependent electric field. It requires a nonzero electric field along the direction perpendicular to magnetization.  (b)-(d) Schematics of five different magnon SPGE currents: (b) the Drude contribution is sorely dominated by the nonequilibrium distribution (the dashed (solid) lines represent the distribution with (without) electric field); (c) the BCD magnon spin current which is determined by the anomalous velocity and the nonequilibrium distribution; (d) the  injection, shift, and rectification spin currents give a combined effect of the dipole transition and group velocity, positional shift, and nonequilibrium distribution.}\label{fig1}
\end{figure}

There are two ways for the electric field entering into the Hamiltonian. Firstly, introducing an ``electric" vector potential $\bm{A}^{E} \equiv \frac{1}{c}\bm{E}(t)\times \hat{\bm{e}}_z$ \cite{ PhysRevLett.90.167204, PhysRevB.95.125429, Owerre_2017, PhysRevB.100.014421} (see Appendix. \ref{appb}), we have
\begin{equation}
\mathcal{H}(\bm{k})= \mathcal{H}_{0}\left(\bm{k} +  \frac{g\mu_{B}}{ c}\bm{A}^{E}\right)
\end{equation}
via the minimal coupling scheme and making use of the Peierls substitution.
Secondly, introducing an effective "electric field" $\tilde{\bm{E}}(t)=-\partial \bm{A}^E /\partial t$, the Hamiltonian is modified in the dipole interaction scheme
(see Appendix. \ref{appc})
\begin{equation}\label{hami-e}
\mathcal{H}(t)=\mathcal{H}_0\left(\bm{k}\right)+\frac{g\mu_B}{c}\tilde{\bm{E}}(t)\cdot \bm{r}.
\end{equation}
In Eq. (\ref{hami-e}), the second term is just the effective dipole interaction.

 Recently, it is recognized that the operator $\bm{r}$ in the crystal lattice is linked to the Berry curvature of Bloch bands of magnons in the momentum space. Thus, a covariant derivative operator rather than a usual partial over the momentum to $\bm{r}$ must be introduced, which is given by \cite{PhysRevB.91.235320, PhysRevB.96.035431}
\begin{equation}\label{cov-de}
\begin{aligned}
(\bm{D}_{\bm{k}}\mathcal{O}_{\bm{k}})_{mn}=\nabla_{\bm{k}}\mathcal{O}(\bm{k})_{mn}-i[\bm{\mathcal{A}}_{\bm{k}},{\mathcal{O}}(\bm{k})]_{mn},
\end{aligned}
\end{equation}
where $\bm{\mathcal{A}}_{\bm{k}}$ is the Berry connection whose matrix elements are $\bm{\mathcal{A}}_{\bm{k}mn}=i\langle u_{\bm{k}m}\ket{\bm{\nabla}_{\bm{k}}u_{\bm{k}n}}$, with $\ket{u_{\bm{k}n}}$ the periodic part of the Bloch functions, and $\mathcal{O}$ is an operator. Noting that the definition of covariant derivative operator in \Eq{cov-de} applies to general magnetic ground state when the magnon Hamiltonian takes the unified transformation $\mathcal{H}(\bm{k})\rightarrow \mathcal{H}(\bm{k}-\frac{g\mu_b}{\hbar c}\bm{A}^E)$ in the presence of AC phase. The perturbed  Hamiltonian is formally the same as  the electron Hamiltonian in the presence of electric field (though the vector potential and the coefficient in front are different).

The effect of electric field is thus incorporated in von Neumann equation (see Appendix. \ref{appe}) as the optical driving term ($\mathcal{D}_{\text{opt}}$ term)
\begin{equation}
\left(\partial_t+i\varepsilon_{\bm{k}mn}/\hbar\right)\rho_{\bm{k}mn}(t)= \mathcal{D}_{\text{opt}} [\rho(t)]_{\bm{k}mn},
\label{iterativeRhon}
\end{equation}
where $\varepsilon_{\bm{k}mn}= \varepsilon_{\bm{k}m}-\varepsilon_{\bm{k}n}$, $\rho_{\bm{k}mn}(t)\equiv \langle \Phi_{ \bm{k}m}^{\dagger}\Phi_{\bm{k}n} \rangle$ is the reduced density matrix  in band space given by the average of the product of a creation and a destruction operator in Bloch states, and
\begin{equation}
\mathcal{D}_{\text{opt}}[\mathcal{O}] =  \frac{g\mu_B }{\hbar c^2}\bm{E}(t)\times \bm{e}_z \cdot \bm{D}_{\bm{k}}[\mathcal{O}].
\end{equation}
The von Neumann equation can be solved by expanding $\rho = \sum_{n=0}\rho^{(n)}$,
where the zero-order one  is the Bose-Einstein distribution  $\rho^{(0)}_{\bm{k}mn}=g_{\bm{k}m}\delta_{mn}$. The recursion equations for the reduced density matrices are obtained as
\begin{equation}\label{recurs}
\begin{aligned}
\rho^{(n+1)}_{mn}(\omega)=& d_{mn}(\omega)\int \frac{d\omega_1}{2\pi}\omega_1  \mathcal{D}_{\text{opt}}[{\rho}^{(n)}(\omega-\omega_1)]_{mn},
\end{aligned}
\end{equation}
with  $d_{mn}(\omega)=1/(\hbar\omega + i0^{+}-\varepsilon_{mn})$.

\begin{table*}[htbp]
	\renewcommand\arraystretch{2}
	\caption{Different terms leading to the magnon SPGE (see \Eq{m-spge-chi}).  The spin photoconductivities are evaluated in terms of the following gauge invariant quantities:  group velocity $v_{m}^{\alpha}=\frac{1}{\hbar}\partial^{\alpha}\varepsilon_{m}$, where $\partial^{\alpha}=\partial/\partial k_{\alpha}$;   velocity difference $\Delta_{mn}^{\alpha}=v_{m}^{\alpha} - v_{n}^{\alpha}$;  Berry curvature $\Omega_{m}^{\tau} = \epsilon^{\alpha\gamma\tau}\partial^{\gamma}\mathcal{A}_{nn}^{\tau}$;  band-resolved quantum metric (Berry curvature) $G_{mn}^{\beta\gamma} =\{ \mathcal{A}_{  mn}^{\beta}, \mathcal{A}_{  nm}^{\gamma }\}/2$ ($\Omega_{mn}^{\beta\gamma} =i[ \mathcal{A}_{  mn}^{\beta}, \mathcal{A}_{  nm}^{\gamma }]$); shift vector $R_{mn}^{\alpha \alpha_1} = \mathcal{A}_{mm}^{\alpha} - \mathcal{A}_{nn}^{\alpha} - \partial^{\alpha} \arg \mathcal{A}_{mn}^{\alpha_1}$; chiral shift vector
$R_{mn}^{\alpha, \pm} = \mathcal{A}_{mm}^{\alpha} - \mathcal{A}_{nn}^{\alpha} - \partial^{\alpha} \arg \mathcal{A}_{mn}^{\pm}$ with the Berry connection in circular representation $\mathcal{A}_{mn}^{\pm} = \frac{1}{\sqrt{2}}(\mathcal{A}_{mn}^{x}\pm i\mathcal{A}_{mn}^{y})$. $\varsigma = \alpha\Omega =1/\tau$ where $\alpha$ is the Gilbert damping constant and $\tau$ is the relaxation time. The conductivities can be written as $\eta^{\alpha\alpha_1\alpha_2}=(\nu_c/2)\int [d\bm{k}] \mathcal{I}^{\alpha\alpha_1\alpha_2}$ and $\kappa^{\alpha\alpha_1\alpha_2}=(\nu_c/2)\int [d\bm{k}] \mathcal{K}^{\alpha\alpha_1\alpha_2}$ with $\mathcal{I}$ and $\mathcal{K}$ being the integrand, $\nu_c$ being the constant $g^2 \mu_{B}^2/(\hbar^2 c^4)$. The symbols $\updownarrow$ and $\circlearrowleft$ denote photocurrents induced by LP and CP light, respectively. Note that the LP responses include a factor $\varsigma = 1$ for $\alpha_1 = \alpha_2$ and $\varsigma = -1$ for $\alpha_1 \neq \alpha_2$.  A phenomenological scattering rate $\Gamma$ is introduced for the injection current. }\label{tab1}
	\begin{tabular*}{17.5cm}{@{\extracolsep{\fill}}p{0.8cm}ccccc}
		\hline\hline
       Current  & Spin photoconductivity & $\mathcal{T}^\prime$ & $\mathcal{P}\mathcal{T}^\prime$ & Physical origin \\ \hline
       $\text{Durde} {\color{blue}\updownarrow}$ & $\mathcal{I}_{\text{D}}^{\alpha\alpha_1\alpha_2} = \varsigma (2/\hbar) \sum_m   v_{m}^{\alpha}\partial^{\alpha_1}\partial^{\alpha_2}g_{  m}$ & $\times$ & $\checkmark$ & \makecell[c]{ Nonequilibrium distribution }\\
       $\text{BCD} {\color{red}\circlearrowleft}$ & $\mathcal{K}_{\text{BCD}}^{\alpha\alpha_1\alpha_2} = (\Omega/\hbar) \sum_{m} (\epsilon^{\alpha\alpha_2\tau}\partial^{\alpha_1} - \epsilon^{\alpha\alpha_1\tau}\partial^{\alpha_2} )\Omega_{m}^{\tau}g_{m}$ & $\checkmark$ & $\times$ & \makecell[c]{Anomalous velocity \\+ nonequilibrium distribution } \\
       $\text{Injection} {\color{blue}\updownarrow}$ & $\mathcal{I}_{\text{Inj}}^{\alpha\alpha_1\alpha_2} = -\varsigma {4\hbar}\sum_{m , n}\frac{\Omega^2}{(\hbar\Omega -\varepsilon_{mn})^2 +\varsigma^2}  \Delta_{mn}^{\alpha} G_{mn}^{\alpha_1\alpha_2}g_{mn}$ & $\times$ & $\checkmark$ & \makecell[c]{Velocity difference\\ + dipole transition }\\
       $\text{Injection}{\color{red}\circlearrowleft}$  & $\mathcal{K}_{\text{Inj}}^{\alpha\alpha_1\alpha_2} = - {2\pi\hbar}\sum_{m , n}\frac{\Omega^2}{(\hbar\Omega -\varepsilon_{mn})^2 +\varsigma^2} \Delta_{mn}^{\alpha} \Omega_{mn}^{\alpha_1\alpha_2} g_{mn}$ & $\checkmark$ & $\times$ & \makecell[c]{Velocity difference \\ + dipole transition} \\
       $\text{Shift}{\color{blue}\updownarrow}$ & $\mathcal{I}_{\text{Sh}}^{\alpha\alpha_1\alpha_2} = \varsigma2\pi \sum_{m, n}\Omega^2 \delta_{\hbar\Omega -\varepsilon_{mn}}R_{mn}^{\alpha\alpha_1} G_{mn}^{\alpha_1\alpha_1}g_{mn}$ & $\checkmark$ & $\times$ & \makecell[c]{Position shift\\ + dipole transition } \\
       $\text{Shift}{\color{red}\circlearrowleft}$ & $\quad \mathcal{K}_{\text{Sh}}^{\alpha\alpha_1\alpha_2} = -4\pi\sum_{m, n} \Omega^2\delta_{\hbar\Omega -\varepsilon_{mn}}(R_{mn}^{\alpha, +}\lvert\mathcal{A}_{nm}^{+}\rvert^2 -R_{mn}^{\alpha, -}\lvert\mathcal{A}_{nm}^{-}\rvert^2) g_{mn}$ & $\times$ & $\checkmark$ & \makecell[c]{Position shift\\ + dipole transition }\\
       $\text{Rectification}{\color{blue}\updownarrow} $ & $\mathcal{I}_{\text{Rec}}^{\alpha\alpha_1\alpha_2} = \varsigma 4 \sum_{m, n}\frac{\Omega^2 }{\hbar\Omega - \varepsilon_{mn}}G_{mn}^{\alpha_1\alpha_2}\partial^{\alpha}g_{mn}$ & $\times$ & $\checkmark$ & \makecell[c]{ Dipole transition\\ +nonequilibrium distribution }\\
       $\text{Rectification} {\color{red}\circlearrowleft}$ & $\mathcal{K}_{\text{Rec}}^{\alpha\alpha_1\alpha_2} = -\sum_{m, n} \frac{\Omega^2}{\hbar\Omega -\varepsilon_{mn}}\Omega_{mn}^{\alpha_1\alpha_2}\partial^{\alpha}g_{mn}$ & $\checkmark$ & $\times$ & \makecell[c]{ Dipole transition\\ +nonequilibrium distribution }\\ \hline
	\end{tabular*}
\end{table*}

\subsection{Magnon spin current}

Due to the conservation of the $z$ component of the total spin, the local magnon spin density $n(\bm{r}_i)=\hbar\sum_{m}\bm{z}_m a_{i,m}^{\dagger}a_{i,m}$ satisfying the continuity equation $\partial n_{\bm{q}} /\partial t + i\bm{q}\cdot \bm{J}_{s} = 0$ ($n_{\bm{q}}$ is Fourier component of $n(\bm{r}_i)$) in the long-wavelength limit. The magnon spin current operator is found as (see Appendix. \ref{appd})
\begin{equation}
\hat{\bm{J}}_s =\hbar\sum_{\bm{k}}\Phi_{\bm{k}}^\dagger \left( \partial_{\bm{k}}\mathcal{H}_{\bm{k}}-i[\bm{\mathcal{A}}_{\bm{k}},{ \mathcal{H}_{\bm{k}}}]\right)\Phi_{\bm{k}}.
\end{equation}
Since the interaction with electric field has been incorporated into the magnon Hamiltonian via an effective gauge potential caused by the AC phase,  the nonlinear spin photocurrent can be handled with standard perturbation theory now.
The $n$th-order magnon spin current is calculated via the formula $J^{\alpha,(n)}_{s}=\int [d\bm{k}]\text{Tr}[\rho^{(n)}\hat{J}_{s}^{\alpha}]$.


\subsection{Magnon spin photoconductivity tensors}
It is instructive to consider the magnon SPGE in ordered  ferromagnetic  insulators, 
which is a dc spin photocurrent response
\begin{equation}
J_s = \chi^{\alpha\alpha_1\alpha_2}E^{\alpha_1}(\Omega)E^{\alpha_2}(-\Omega),
\end{equation}
with $\Omega$ being the frequency of the light.
The photocurrent response is classified as the linearly polarized LP and CP light-induced currents, the LP (CP)-photocurrent is given by the real symmetric (imaginary antisymmetric) component of the photoconductivity tensor $\eta^{\alpha\alpha_1 \alpha_2} =\frac{1}{2}\text{Re}(\chi^{\alpha\alpha_1 \alpha_2} + \chi^{\alpha\alpha_2 \alpha_1})$,
$[\kappa^{\alpha\alpha_1 \alpha_2} = \frac{1}{2}\text{Im}(\chi^{\alpha\alpha_1 \alpha_2}- \chi^{\alpha\alpha_2 \alpha_1})]$.
We show that $\chi^{\alpha\alpha_1\alpha_2} \propto \epsilon^{\alpha_1 z \beta}\epsilon^{\alpha_2 z \gamma}$ ($\epsilon^{\alpha_2 z \gamma}$ is the Levi-Civita symbol). That is, the electric field is restricted into the plane perpendicular to the axis of magnetization, attributed to the orthogonality between the electric field and the magnon magnetic moment, as illustrated by \Eq{ac-phase} and a schematic in \Fig{fig1}(a).  While for the charge photocurrent there is no such constraint,  making it as a prominent character. 
Leaving the details of the calculation in Appendix. \ref{appe}, we obtain
\begin{equation}\label{m-spge-chi}
\chi^{\alpha\alpha_1\alpha_2} = \chi^{\alpha\alpha_1\alpha_2}_{\text{D}}+\chi^{\alpha\alpha_1\alpha_2}_{\text{BCD}}+\chi^{\alpha\alpha_1\alpha_2}_{\text{Inj}}+\chi^{\alpha\alpha_1\alpha_2}_{\text{Sh}}+\chi^{\alpha\alpha_1\alpha_2}_{\text{Rec}},
\end{equation}
with the functional forms of all contributions tabulated in Table \ref{tab1}.
We denote different contributions as follows: Drude ($\chi^{\alpha\alpha_1\alpha_2}_{\text{D}}$), Berry curvature dipole (BCD) ($\chi^{\alpha\alpha_1\alpha_2}_{\text{BCD}}$), injection ($\chi^{\alpha\alpha_1\alpha_2}_{\text{Inj}}$), shift ($\chi^{\alpha\alpha_1\alpha_2}_{\text{Sh}}$), and the rectification current ($\chi^{\alpha\alpha_1\alpha_2}_{\text{Rec}}$). Noting that in Ref. \cite{PhysRevB.100.224411} a magnon shift spin current is proposed which is driven by the magnetic field component of the  electromagnetic wave. In contrast, the magnon shift current  here manifests as the response of the electric field.

The physical meaning of each contribution in \Eq{m-spge-chi} is schematically shown in \Fig{fig1}(b)-(d). 
The  Drude-spin-current (DSC)  arises from the second-order correction to the distribution function and the band gradient velocity, hence, it manifests as a pure intraband effect. Just as the nonlinear Drude contribution for charge spin current \cite{PhysRevLett.86.4358, PhysRevB.95.224430}, the nonlinear Drude current for magnon has an additional symmetry for swapping the current and field direction indices. As a consequence, it is allowed under LP light but is forbidden under the CP light. 

Similar to the BCD-resulted nonlinear charge current, the BCD-spin-current (BSC) for magnon is originated from a dipole moment of the magnon Berry curvature in  momentum space, and it is classified as a CP response. However, distinguished from   the charge counterpart, the BSC for magnon scales with the frequency of light and hence vanishes as $\Omega$ approaches to zero. It is worth noting that the magnon BSC driven by temperature gradient has been proposed \cite{PhysRevResearch.4.013186}. While here we show that there is a BCD contribution as an electric field response, which has not been reported yet. 

For the charge case, the shift current due to an instantaneous shift in the charge distribution upon absorption of light, and the injection current due to the injection of a carrier distribution that is asymmetric in momentum space have been extensively studied which are considered the main bulk photovoltaic effect \cite{ kirk2017reconsidering, PhysRevX.11.011001,10.1063/5.0101513,  PRXEnergy.2.013006}.  In analogy, we found that the  shift-spin-current (SSC) for magnon can be described as a coherent response associated with the real-space shift of an magnon induced by a dipole-mediated vertical inter-band transition.  For LP light response, the shift vector $R_{mn}^{\alpha, \alpha_1}$ recovers the will-known expression as the charge case \cite{doi:10.1126/sciadv.1501524, PhysRevB.96.075421}, which implies the wave-packet shift of the excited magnons along the $\alpha$ direction through the interband transition $m\leftrightarrow n$. Notably, the chiral shift vector  $R_{mn}^{\alpha, \pm}$ characterizes the handedness of the dipole-transition amplitude denoted by $\mathcal{A}_{mn}^{\pm}$ excited by CP light, resulting in a gyration MSPC. Similar gyration charge current is recently discovered in magnetically parity-violating system \cite{PhysRevX.11.011001}.

The injection-spin-current (ISC) arises from the velocity difference ($\Delta_{mn}$) during the interband transition. It is known that the charge injection current  scales linearly with respect to the relaxation time $\tau$ and is expected to exhibit a large current response when the electron lifetime is long \cite{10.1063/5.0101513}, and vanishes with strong scattering as $\tau\rightarrow 0$. As a contrast, we found that the ISC in our proposal remains finite within the limit $\tau\rightarrow 0$.   It is worth noting that the ISC proposed are found as both CP and LP responses. 

Finally, the rectification spin currents (RSC) as both LP and CP responses are exposed in this work, which are proportional to the derivatives of the distributions, and can be regarded as an analog of the ``intrinsic Fermi surface effect" in electronic system \cite{PhysRevX.11.011001}. Interestingly, from Table \ref{tab1}, it is found that the LP responses (apart from the DSC since it is not geometry related) are determined by the \textit{band-resolved quantum metric} while the CP responses are determined by the \textit{Berry curvature}. Our quantum kinetic theory based treatment of the electric field interaction via the effective dipole interaction provides a complete picture of the magnon SPGE, as summarized in \Fig{fig1}(b)-(d).

\subsection{Symmetry characters}

Since the present method and developed theory are firstly proposed, a symmetry analysis is quite useful to outline the scope of this study.
%
For a magnet with collinear magnetic order, it  is invariant under the combined symmetry operation of the time-reversal $\mathcal{T}$ and a $\pi$ spin rotation about the axis perpendicular to the plane of the magnetic order. This is called the effective time-reversal symmetry $\mathcal{T}^\prime$ \cite{PhysRevLett.112.017205, PhysRevB.95.094406}. A general Hamiltonian can be transformed with the effective TRS operation $\mathcal{T}^\prime$ as
\begin{equation}\label{etrs}
\mathcal{T}^\prime \mathcal{H}_{\bm{k}}(\mathcal{T}^\prime)^{-1} = \mathcal{H}_{-\bm{k}}=\mathcal{H}_{\bm{k}}^{*}.
\end{equation}
It gives rise to 
\begin{equation}\label{eigen1}
U_{\bm{k}}^{\dagger}\mathcal{H}_{\bm{k}}U_{\bm{k}} = U_{\bm{k}}^{\dagger}\mathcal{H}_{-\bm{k}}^* U_{\bm{k}} = \mathcal{E}_{\bm{k}}.
\end{equation}
Noting that the eigenvalues are real, we have 
\begin{equation}\label{eigen2}
U_{-\bm{k}}^{T}\mathcal{H}_{\bm{k}}U_{-\bm{k}}^* = (U_{-\bm{k}}^*)^{\dagger}\mathcal{H}_{\bm{k}} U_{-\bm{k}} = \mathcal{E}_{-\bm{k}}.
\end{equation}
The upperscript $T$ stands for transpose. 
\Eq{eigen1} and \Eq{eigen2} indicate that $U_{\bm{k}}^{\dagger}$ and $U_{-\bm{k}}^*$  differ only by a phase factor matrix, and share the same eigenvalue equation with $\mathcal{E}_{\bm{k}}=\mathcal{E}_{-\bm{k}}$.
The element of the Berry connection matrix is
\begin{equation}
\mathcal{A}_{\bm{k}mn}^{\alpha} = i\sum_{p}(U^{\dagger}_{\bm{k}})_{mp} \frac{\partial (U_{\bm{k}})_{pn}}{\partial k_{\alpha}}=i\sum_{p}U_{\bm{k}pm}^{*} \frac{\partial U_{\bm{k}pn}}{\partial k_{\alpha}}.
\notag
\end{equation}
The $\mathcal{T}^\prime$ symmetry gives a constraint on the Berry connection
$\mathcal{A}_{\bm{k}mn}^{\alpha} =\mathcal{A}_{-\bm{k}nm}^{\alpha}$, and
the Berry curvature then satisfies
$\Omega_{\bm{k}m}^{\tau} =-\Omega_{-\bm{k}m}^{\tau}$.
Similarly, with $\mathcal{T}^\prime$, the band-resolved Berry curvature (Eq. (S113)), band-resolved quantum metric and the shift vector (Eq. (S113)) satisfy
\begin{align}
\Omega_{\bm{k}mn}^{\alpha\beta} =& -\Omega_{-\bm{k}mn}^{\alpha\beta}, \label{interbc-trnas}\\
G_{\bm{k}mn}^{\alpha\beta} = &  G_{-\bm{k}mn}^{\alpha\beta}. \label{interqm-trans}\\
R_{\bm{k}mn}^{\alpha\beta} = & R_{-\bm{k}mn}^{\alpha\beta}. \label{shift-vec-trans}
\end{align}

\begin{table*}
\tabcolsep=0.5mm
	\renewcommand\arraystretch{2}
	\caption{Classification on the spin photoconductivities the Drude, BCD, injection, shift, and rectification spin current for LP and CP light by the point-group operation and the effective time-reversal $\mathcal{T}^\prime$.  The allowed (forbidden) conductivities are indicated by $\checkmark$ ($\times$). }
	\resizebox{\textwidth}{36mm}{\begin{tabular*}{19.5cm}{@{\extracolsep{\fill}}p{0.7cm}ccccccccccccccccccc}
		\hline\hline
         & $\mathcal{P}$ & $C_{2}^{y}$ & $C_{2}^{z}$ & $\mathcal{P}C_{2}^{y}$ & $\mathcal{P}C_{2}^{z}$ & $C_{3}^{z}$ & $C_{4}^{z}$ & $\mathcal{P}C_{4}^{z}$ & $C_{4}^{z}\sigma_v$ & $\mathcal{T}^\prime$ & $\mathcal{P}\mathcal{T}^\prime$ & $C_{2}^{y}\mathcal{T}^\prime$ & $C_{2}^{z}\mathcal{T}^\prime$ & $\mathcal{P}C_{2}^{y}\mathcal{T}^\prime$ & $\mathcal{P}C_{2}^{z}\mathcal{T}^\prime$ & $C_{3}^{z}\mathcal{T}^\prime$ & $C_{4}^{z}\mathcal{T}^\prime$ & $\mathcal{P}C_{4}^{z}\mathcal{T}^\prime$ & $C_{4}^{z}\sigma_v\mathcal{T}^\prime$
          \\
		\hline
        $\eta_{\text{D}}^{\alpha\alpha_1 \alpha_2}$
        &  $\times$ &  $\checkmark$ &  $\times$ & $\checkmark$ &   $\checkmark$ & $\checkmark$&  $\times$& $\times$ &$\checkmark$&$\times$& $\checkmark$& $\checkmark$ & $\checkmark$ &$\checkmark$ &$\times$ & $\times$& $\times$& $\times$& $\times$
        \\
        \hline
	  $\kappa_{\text{BCD}}^{\alpha\alpha_1 \alpha_2}$
	  &  $\times$ & $\checkmark$  &  $\times$ & $\checkmark$ &  $\checkmark$ &$\times$&$\times$& $\times$ & $\checkmark$ & $\checkmark$& $\times$& $\checkmark$ & $\checkmark$ &$\checkmark$ &$\checkmark$ & $\times$& $\times$& $\times$& $\times$\\
	  \hline
	  $\eta_{\text{Inj}}^{\alpha\alpha_1 \alpha_2}$
        &  $\times$ &  $\checkmark$ &  $\times$ & $\checkmark$ &   $\checkmark$ &$\checkmark$&$\times$& $\times$ & $\checkmark$ & $\times$& $\checkmark$ & $\checkmark$ & $\checkmark$ &$\checkmark$ &$\times$& $\times$& $\times$& $\times$& $\times$
        \\ \hline
        $\kappa_{\text{Inj}}^{\alpha\alpha_1 \alpha_2}$
	  &  $\times$ & $\checkmark$  &  $\times$ & $\checkmark$ &  $\checkmark$&$\times$&$\times$& $\times$  &$\checkmark$& $\checkmark$& $\times$ & $\checkmark$ &$\checkmark$ &$\checkmark$ &$\times$& $\times$& $\times$& $\times$& $\times$\\ \hline
	  $\eta_{\text{Sh}}^{\alpha\alpha_1 \alpha_2}$
        &  $\times$ &  $\checkmark$ &  $\times$ & $\checkmark$ &   $\checkmark$ &$\checkmark$ & $\times$& $\times$& $\checkmark$& $\checkmark$&$\times$& $\checkmark$ & $\checkmark$ &$\checkmark$ &$\checkmark$& $\times$& $\times$& $\times$& $\times$
        \\
        \hline
	  $\kappa_{\text{Sh}}^{\alpha\alpha_1 \alpha_2}$
	  &  $\times$ & $\checkmark$  &  $\times$ & $\checkmark$ &  $\checkmark$ &$\times$&$\times$& $\times$  &$\checkmark$& $\times$& $\checkmark$& $\checkmark$ &$\checkmark$ &$\checkmark$ &$\checkmark$& $\times$& $\times$& $\times$& $\times$\\
	  \hline
	  $\eta_{\text{Rec}}^{\alpha\alpha_1 \alpha_2}$
        &  $\times$ &  $\checkmark$ &  $\times$ & $\checkmark$ &   $\checkmark$ &$\checkmark$ & $\times$& $\times$& $\checkmark$ & $\times$&$\checkmark$& $\checkmark$ &$\checkmark$ &$\checkmark$ &$\times$& $\times$& $\times$& $\times$& $\times$
        \\ \hline
        $\kappa_{\text{Rec}}^{\alpha\alpha_1 \alpha_2}$
	  &  $\times$ & $\checkmark$  &  $\times$ & $\checkmark$ &  $\checkmark$&$\times$&$\times$& $\times$  &$\checkmark$& $\checkmark$& $\times$ & $\checkmark$ &$\checkmark$ &$\checkmark$ &$\times$& $\times$& $\times$& $\times$& $\times$\\
		\hline\hline
	\end{tabular*}}
\label{tab2}
\end{table*}

Let's consider the magnon spin photoconductivities. $\eta_{s,\text{Drude}}^{\alpha\alpha_1 \alpha_2}$ is $\mathcal{T}^\prime$-odd, indicating that they are finite only when $\mathcal{T}^\prime$ are broken.
%
The integrand of $\kappa_{s,\text{BCD}}^{\alpha\alpha_1 \alpha_2}$ transforms as
\begin{equation}
\partial^{\alpha_1}\Omega_{\bm{k}m}^{\tau} = \partial^{\alpha_1}\Omega_{-\bm{k}m}^{\tau}.
\end{equation}
It renders that $\kappa_{s,\text{BCD}}^{\alpha\alpha_1 \alpha_2}$ is $\mathcal{T}^\prime$-even.
For the LP injection current $\eta_{s,\text{inj}}^{\alpha\alpha_1 \alpha_2}$, we have
\begin{equation}
\Delta_{\bm{k}mp}^{\alpha}G_{\bm{k}mp}^{\beta\gamma}g_{\bm{k}mp} = -\Delta_{-\bm{k}mp}^{\alpha}G_{-\bm{k}mp}^{\beta\gamma}g_{-\bm{k}mp},
\end{equation}
leading to that the LP ISC is odd under $\mathcal{T}^\prime$.
Similarly, the integrand of CP ISC $\kappa_{s,\text{inj}}^{\alpha\alpha_1 \alpha_2}$ satisfies
\begin{equation}
\Delta_{\bm{k}mp}^{\alpha}\Omega_{\bm{k}mp}^{\beta\gamma}g_{\bm{k}mp} = \Delta_{-\bm{k}mp}^{\alpha}\Omega_{-\bm{k}mp}^{\beta\gamma}g_{-\bm{k}mp},
\end{equation}
 i.e., the CP ISC is $\mathcal{T}^\prime$-even.
For the LP SSC  $\eta_{s,\text{shift}}^{\alpha\alpha_1 \alpha_1}$,  the integrand satisfies
\begin{equation}
 R_{\bm{k}mn}^{\alpha_1\alpha_1} G_{\bm{k}mn}^{\alpha_1\alpha_1} g_{\bm{k}mn}=R_{-\bm{k}mn}^{\alpha_1\alpha_1} G_{-\bm{k}mn}^{\alpha_1\alpha_1} g_{-\bm{k}mn},
\end{equation}
suggesting that  $\eta_{s,\text{shift}}^{\alpha\alpha_1 \alpha_2}$ is $\mathcal{T}^\prime$-even.
The CP SSC $\kappa_{s,\text{shift}}^{\alpha\alpha_1 \alpha_2}$ is justified to be $\mathcal{T}^\prime$-odd in a similar way.
In analogy, the  LP RSC $\eta_{s,\text{rect}}^{\alpha\alpha_1 \alpha_1}$ is  $\mathcal{T}^\prime$-odd and the CP RSC $\kappa_{s,\text{rect}}^{\alpha\alpha_1 \alpha_2}$ is  $\mathcal{T}^\prime$-even.

Now we consider the point-group symmetry transformations. With a point-group symmetry $\mathcal{M}$, the eigenvalues satisfy $\varepsilon_{n\bm{k}} = \varepsilon_{n\mathcal{M}^{-1}\bm{k}}$, and the Berry connection transforms as
\begin{equation}
\mathcal{A}_{\bm{k}mn}^{\alpha} = \mathcal{M}_{\alpha\beta}\mathcal{A}_{\mathcal{M}^{-1}\bm{k}mn}^{\beta}.
\end{equation}
The derivative operation transforms as
\begin{equation}
\frac{\partial g_{\bm{k}m}}{\partial \bm{k}_{\alpha}}=\mathcal{M}_{\alpha\beta}\frac{\partial g_{\mathcal{M}^{-1}\bm{k}m}}{\partial \bm{k}_{\beta}}.
\label{distr-trans-2}
\end{equation}
From these properties, for a general conductivity tensor, one can find that
\begin{equation}
\chi^{\alpha^\prime \alpha_{1}^\prime \alpha_{2}^\prime} = \mathcal{M}^{\alpha^\prime \alpha}\mathcal{M}^{\alpha_{1}^\prime \alpha_1 }\mathcal{M}^{\alpha_{2}^\prime\alpha_2 }\chi^{\alpha\alpha_1\alpha_2}.
\end{equation}
For a CP response tensor, it is antisymmetric by permuting the last two indices, and  it is convenient to transform it to an equivalent rank-2 pseudotensor
\begin{equation}
\varpi^{\alpha\beta} = \epsilon^{\beta\alpha_1\alpha_2}\kappa^{\alpha\alpha_1\alpha_2}/2,
\end{equation}
and the operation of $\mathcal{M}$ yields
\begin{equation}
\varpi^{\alpha^\prime\beta^\prime} = \det{(\mathcal{M})}\mathcal{M}^{\alpha^\prime \alpha}\mathcal{M}^{\beta^\prime \beta}\varpi^{\alpha\beta}.
\end{equation}
The constraints on the in-plane magnon spin photoconductivity tensors from the point-group symmetries and the effective time-reversal $\mathcal{T}^\prime$ are listed in Table \ref{tab2}. It is seen that the existence of the magnon spin current requires to break either  $\mathcal{T}^{\prime}$ or $\mathcal{P}\mathcal{T}^\prime$ symmetry. For example, the Drude contribution is forbidden by $\mathcal{T}^\prime$, while the Berry curvature dipole contribution is forbidden by $\mathcal{P}\mathcal{T}^\prime$. Further constraints on the magnon spin currents are introduced by the point group symmetries. For example, the $C_{3z}$ symmetry forces all the circularly-polarized-light responses of the magnon spin current to vanish.

\section{Results}

\subsection{Application to breathing kagome ferromagnet}

\begin{figure}[tb]
\centering
\includegraphics [width=1\columnwidth]{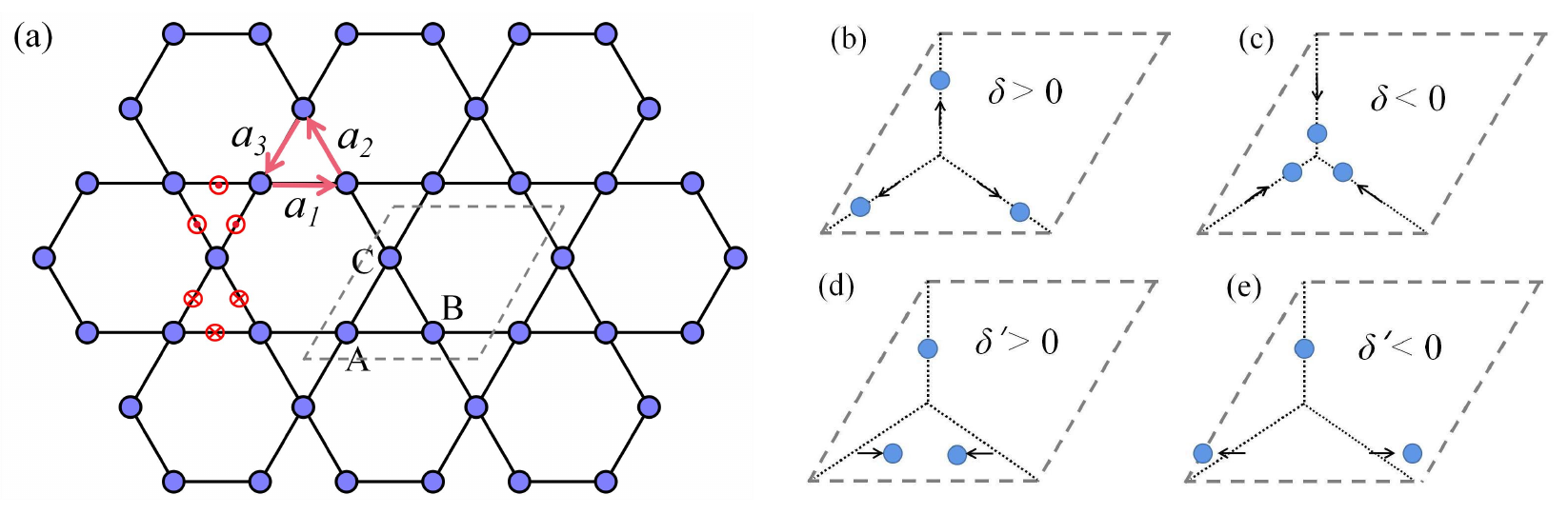}
\caption{(a) Schematics for the kagome ferromagnet with lattice constant $a$. The NN vectors are labeled by $a_i$. Sublattices A, B, and C are placed at the corners of the triangles. Dzyaloshinskii-Moriya vectors are aligned normal to the lattice plane, and their directions (up and down) are represented by the symbols $\odot$ and $\otimes$, depending on the chirality of the triangles. The dashed lines represent a unit cell. (b)-(c) Strain is introduced by letting the sublattice further away ($\delta > 0$) or closer ($\delta < 0$).   }\label{fig2}
\end{figure}

The breathing kagome lattice has gained growing interest in the studies of the quantum spin liquid  \cite{essafi2017flat,schaffer2017quantum,ezawa2018higher,bolens2019topological,PhysRevB.104.144422, PhysRevB.105.014404}. 
To illustrate the proposed magnon spin photogalvanic effect,  we study  a breathing kagome-lattice ferromagnets in the absence of inversion symmetry, as shown in \Fig{fig2} (a). 
With a nonuniform strain field, the three sublattices are deformed further away (positive perturbation $\delta > 0$) or getting closer (negative perturbation $\delta < 0$) from their shared corner, as shown in \Fig{fig2} (b) and (c).  The Hamiltonian is
\begin{equation}
H= - \sum_{\langle i,j\rangle}J_{ij}\bm{S}_{i}\cdot \bm{S}_{j} + \sum_{\langle i,j \rangle}\bm{D}_{ij}\cdot \bm{S}_{i}\times \bm{S}_{j}.
\label{hami-model1}
\end{equation}
$J>0$ is the nearest-neighbors (NN) ferromagnetic coupling strength.
The second term is out-of-plane NN Dzyaloshinskii-Moriya interaction (DMI), the Dzyaloshinskii-Moriya (DM) vector is specified as $\bm{D}_{ij} = v_{ij}D\bm{z}$ with $v_{ij}=\pm$,  depending on chirality of the triangles in the kagome lattice.
%
%
%
%
The ferromagnetic exchange interaction and the DMI are given by   \cite{PhysRevB.97.054404,PhysRevB.104.144422}
$J_{\pm} = (1 \mp \sqrt{3}\eta \delta)J$,
$D_{\pm} = (1 \mp \sqrt{3}\eta \delta)D$,
where subscript $+$ ($-$) denotes intracell (intercell) NN couplings, and $\eta$ is a parameter describing the response of the couplings to the displacements of sublattices, $\delta$ is the strain parameter.
The  bosonic Hamiltonian can be derived as $\mathcal{H}_{\bm{k}}= \mathcal{H}_0 - \mathcal{H}_{J}^{NN} - \mathcal{H}_{DM}$, where $\mathcal{H}_0  =4J I_{3\times 3} $. The nearest neighbour magnetic exchange coupling reads
\begin{equation}
\mathcal{H}_{J}^{NN} = \begin{pmatrix}
0 & \gamma_2 & \gamma_{3}^* \\ \gamma_{2}^* & 0 & \gamma_1 \\
\gamma_3 &  \gamma_{1}^* & 0
\end{pmatrix},  \mathcal{H}_{DM} = i \begin{pmatrix}
0 & d_1 & -d_{3}^* \\ -d_{1}^* & 0 & d_2 \\
d_3 &  -d_{2}^* & 0
\end{pmatrix},
\end{equation}
with $\gamma_i = J_{+}e^{i\bm{k}\cdot(1+\delta)\bm{a}_i} + J_{-}e^{-i\bm{k}\cdot(1-\delta)\bm{a}_i}$ and $d_i = D_{+}e^{i\bm{k}\cdot(1+\delta)\bm{a}_i} + D_{-}e^{-i\bm{k}\cdot(1-\delta)\bm{a}_i}$.

\begin{figure}[tb]
\centering
\includegraphics [width=3.3in]{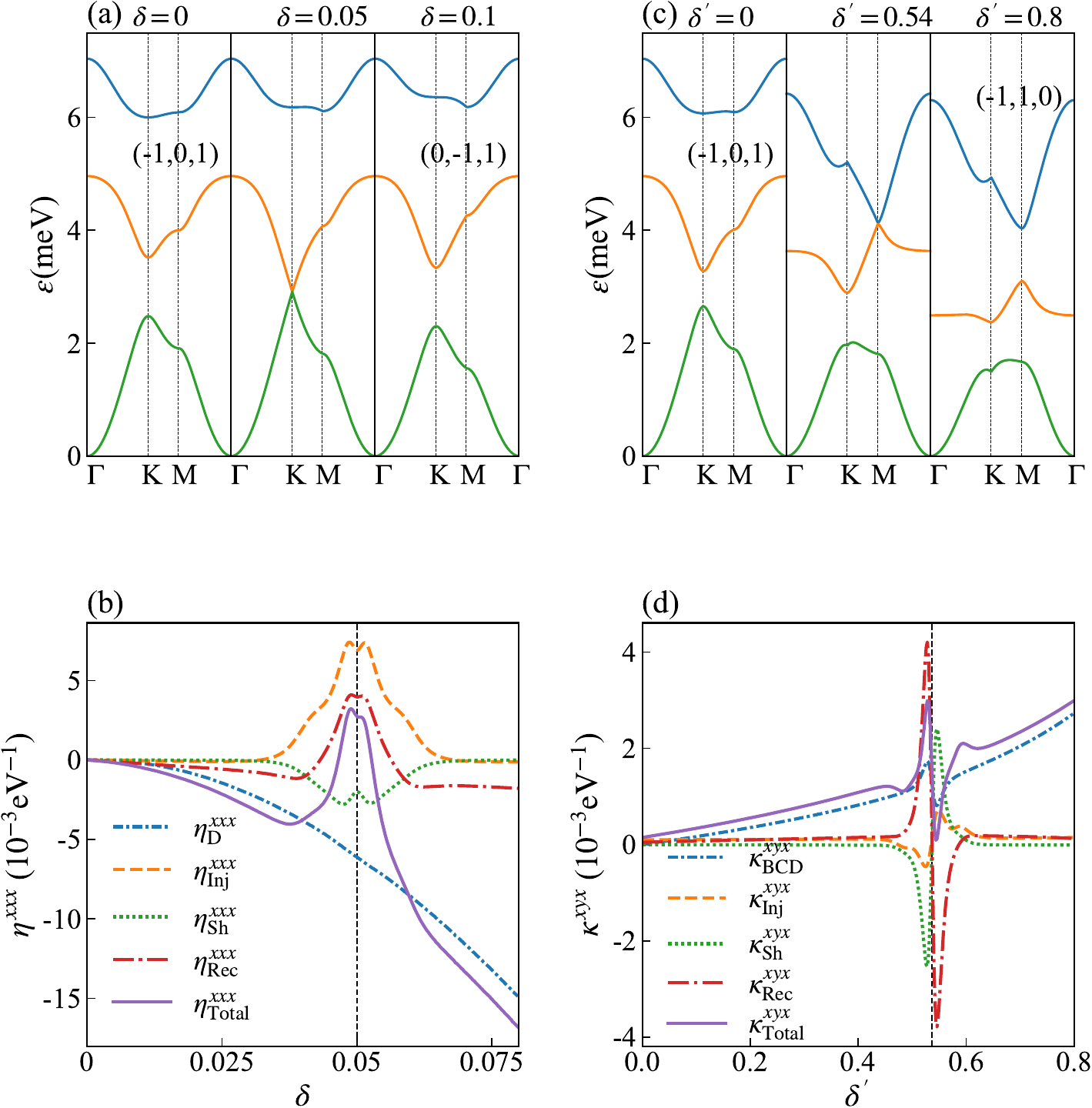}
\caption{(a) Magnon band structures of a breathing kagome-lattice ferromagnet with different lattice deformation $\delta$. Considering that $\delta > 0$ and $\delta < 0$ are equivalent, only cases $\delta > 0$ are plotted. (b) LP photoconductivities as a function of the lattice deformation. The critical point of phase transition is shown as vertical dashed line.  (c) Magnon band structures of a breathing kagome-lattice ferromagnet in the presence of $\delta=0.02$ with different uniaxial strain $\delta^\prime$. The critical point is $\delta^\prime =0.54$.  (d) The CP photoconductivities versus the uniaxial strain $\delta^\prime$, where the critical point of phase transition is shown as vertical dashed line.  Parameters are set as $J=3.405$, $D = 1.02$, $\hbar\Omega = 0.15$, in unit meV. Gilbert damping constant $\alpha$ is estimated as $\alpha = 10^{-2}$,  temperature is $T = 50$ K. }\label{fig3}
\end{figure}

Different topological phases are characterized by sets of Chern numbers $(C_1, C_2, C_3)$ of the lower, middle and upper magnon bulk bands.
A topological phase transition has been discovered by tuning $\delta$ \cite{PhysRevB.97.054404}. \Fig{fig3}(a) shows the magnon bulk bands along the high-symmetry directions ($\Gamma$-$K$-$M$-$\Gamma$) of the Brillouin zone.
For $\delta =0$ the topological phase is $(-1,0,1)$ and for $\delta = 0.1$ the topological phase is $(0,-1,1)$. A topological phase transition occurs at $\delta_c = 0.05$, where  the two  magnon branches cross linearly at the $K$ point.

The DMI breaks the effective $\mathcal{T}^\prime$ symmetry, it also reduces the $C_{6z}$ symmetry  to $C_{3z}$, yet preserves the $C_{2x}$ symmetry.
Then both of the LP and CP responses are forbidden. It is worthy noting that the breathing geometry preserves the $C_{3z}$ symmetry and breaks $C_{2x}$,  hence all the CP responses are forbidden but the LP responses are allowed.
In \Fig{fig3}(b) we show the magnon spin photoconductivities $\eta^{xxx}$ of the allowed injection, rectification, and shift spin current in response of the LP light, varying with the lattice deformation.
For $\delta =0$, the LP responses are zero owing to the $C_{2x}$ symmetry. As $\delta$ increases, finite LP responses appear as required by symmetry. The $(-1,0,1)$ phase and the  $(0,-1,1)$ phase are separated by the critical points of phase transition represented by the dashed black lines.
Remarkably, both of the injection and the rectification  spin current exhibits a peak at the critical point; while the derivative of the shift spin current maximizes at the critical point.
As for the DSC, there is no such drastic change at the critical point, because the DSC is exclusively determined by the group velocities and their derivatives, which is independent to the magnon topology.
Thus it clearly indicates a topological phase transition that is also manifested as a plateau of the total LP photoconductivity at the critical point, which can serve as a prominent indicator to determine the topological phase transition in experiments.

A breathing kagome-lattice ferromagnet preserves the $C_{3z}$ symmetry hence all the CP responses are forbidden. Now we consider an additional uniaxial strain along the $x$-axis breaking the $C_{3z}$ symmetry, for which the spins along the $x$-axis are pushing closer $\delta^\prime >0$ or further $\delta^\prime <0$, as shown in \Fig{fig2}(d)-(e).
The remaining point-group symmetry is $\sigma_x$, for which the CP responses are allowed. In the presence of  $\delta^\prime$, the diagonal component $\mathcal{H}_0$ is modified as  $\mathcal{H}_0  = J\text{diag} (4,2(2+\delta^\prime), 2(2+\delta^\prime))$. For $\mathcal{H}_{J}^{NN}$, only the $(1,2)$ and $(2,1)$ elements are changed to $\gamma_{2}^\prime = (1+\delta^\prime)\gamma_2$ \cite{owerre2018strain}.
By increasing $\delta^\prime$ from zero,  a topological phase transition  occurs at the critical point $\delta_{c}^\prime =\frac{1}{2}[1-(D/J)^2]$, where the upper band and the middle band cross at the M point.
In \Fig{fig3}(c) we show the magnon bands along the high symmetry points for different $\delta'$ but with fixed $\delta = 0.02$ (and $D/J=0.2$).
Similar to Ref. \cite{owerre2018strain}, a topological phase transition occurs when the system is tuned from the $(-1,0,1)$-phase to $(-1,1,0)$-phase  by increasing $\delta^\prime$; while the critical point is found at $\delta_{c}^\prime = 0.54$.
In \Fig{fig3}(d) we show the CP spin conductivities $\kappa^{xyx}_{s}$  as functions of $\delta^\prime$,  where we set the circular polarization of light to be in the $xy$ plane and measure the $z$-direction spin current. One observes that all four kinds of the CP responses change abruptly at $\delta^\prime = \delta_{c}^\prime$, indicating a topological magnon phase transition, which is a direct manifestation of the change of the topological properties such as the Berry curvature and the shift vector in momentum space.

\subsection{Materials realization}
In above we qualitatively analyse the response tensors by symmetry and characterize the behaviour near the  phase transition point.  
Now we consider the magnitude for the magnon spin photoconductivity. The candidate materials can be the three-dimensional (3D) ferromagnetic pyrochlore oxides $\rm{Lu_2 V_2 O_7}$, $\rm{In_2 Mn_2 O_7}$, as well as $\rm{Ho_2 V_2 O_7}$ \cite{onose2010observation},  with a nonuniform strain applied. Instead of considering a “true” 3D lattice, we can realize our idea by treating the system as a stack of noninteracting (or weak-interacting) kagome layers. Here we consider the $\rm{Lu_2 V_2 O_7}$ with nonuniform strain applied. The lattice constant is $a=7.024 \rm{\AA}$ and the Curie temperature is $70$ K \cite{onose2010observation}. The nearest Heisenberg interaction is estimated as $J_1=3.405$ meV  \cite{PhysRevB.89.134409} and we set the temperature to $T=50$ K with a ferromagnetic ground state. From these values and our numerical result in \Fig{fig3}(c) and \Fig{fig3}(d), we obtain that the total  magnon spin photoconductivity under the LP and CP light can reach $10^{-2}$ e/V near the phase transition points. When applying an ac electric field of $E\sim 10^3$ V/m, it leads to a magnon spin current $J_s\sim 10^{-2}$ eV/$\rm{cm^2}$, which is experimentally detectable. The generated magnon spin  current can be detected via the inverse spin Hall effect   \cite{cornelissen2015long}.

Though the AC phase scales with $1/c^2$, the obtained observable magnon spin currents induced by AC phase can be understood as follows.  We can compare the AC-phase-resulted magnon spin currents to electron spin currents in response to electric field. To be concrete, supposing that we consider  a circularly polarized laser (electromagnetic) field with the electric field component $\bm{E}(t)=E_0(\cos{\omega t}, \sin{\omega t},0)$. By use of $\bm{E}(t)=-\partial_t A(t)$,  the magnetic vector potential is written as $\bm{A}(t)=A_0(\sin{\omega t}, -\cos{\omega t},0)$. The AC phase  is given by $\theta_{ij}^{\text{AC}}=\frac{g\mu_{B}}{\hbar c^2}\int_{\bm{r}_i}^{\bm{r}_j}(\bm{E}(t)\times \hat{\bm{e}}_{z})\cdot d\bm{r}$, which can be rewritten as $\theta_{ij}^{\text{AC}}=\frac{g\mu_{B}\omega}{\hbar c^2}\int_{\bm{r}_i}^{\bm{r}_j}(\bm{A}(t+\frac{\pi}{2\omega})\times \hat{\bm{e}}_{z})\cdot d\bm{r}$. As is known, the Aharonov-Bohm phase is $\theta_{ij}^{\text{AB}}=\frac{q}{\hbar }\int_{\bm{r}_i}^{\bm{r}_j}\bm{A}(t)\cdot d\bm{r}$. It renders that for the same vector potential $\bm{A}$ or electric field $\bm{E}$, there is the relation $\theta_{ij}^{\text{AC}}/\theta_{ij}^{\text{AB}} \sim g\mu_B\omega/q c^2$, implying  a considerable AC effect with large $\omega$. For example, a 100THz electric field results to $g\mu_B\omega/q c^2 \sim 10^{-6}$. Consequently,  when increasing the electric field to  three orders  larger (considering that the magnon spin photogalvanic effect is of the second-order response), the magnitude of the magnon spin current can approach to that of the charge spin current case.

 Noting that the intrinsic (without nonuniform strain) breathing kagome ferromagnet has been synthesized in rare-earth based pyrochlore materials like  $\mathrm{Ba_{3}Yb_{2}Zn_{5}5O_{11}}$ \cite{PhysRevLett.116.257204}, as well as  $\mathrm{LiZn_{2}Mo_{3}O_{8}}$ \cite{PhysRevLett.120.227201}. Also, a field-aligned ferromagnetic phase  was experimentally observed in the centrosymmetric breathing kagome lattice  $\mathrm{Gd_{3}Ru_{4}Al_{12}}$  \cite{hirschberger2019skyrmion}. We expect the MSPC we proposed can be observed in these candidate materials.
With the symmetry analysis, this study can be naturally extended to other materials by using the magnetic structure database MAGNDATA \cite{gallego2016magndata}.

\section{Discussion and conclusion}

To conclude, we propose a magnon spin photogalvanic effect and uncover the geometric origin due to an AC phase accumulation. It offers a new mechanism for optical (electric) generation and control of the magnons.  Since the electric component of light is of wide tunable parameter scale and very fast, this method may find a realistic platform in applications.
We further identify five new mechanisms: the Drude, BCD, injection, shift, and the rectification spin current. 
It is found that the linearly-polarized-light responses (except Drude) are determined by the band-resolved quantum metric; while the circularly-polarized-light responses are determined by the Berry curvature. Noting that in general the magnon  does not conserve the spin angular momentum when considering the  processes  such as the dipolar interaction \cite{PhysRevLett.116.146601} and magnon-magnon interaction that involve the pair creation and annihilation of magnons, as well as the scattering with magnetic impurity that disturbs the magnon spin angular momentum \cite{PhysRevLett.110.157203}. 
 Here we assume the situation that the Heisenberg  exchange coupling is strong enough to suppress the dipolar interaction \cite{kittel1963quantum}, the temperature is low and the impurities are sparse  that the magnon-magnon interaction and impurity scattering are negligible \cite{nolting2009quantum}, then we can restrict the study to the circumstance that the magnon preserves the spin angular momentum $\pm \hbar$ when transfers among sites, and we leave the generalization that the magnon does not conserve the spin angular momentum for future study. 
Finally, we have to emphasize that the present calculation is also valid for the case of an ac bias applied to the sample and may be of great interests to experiments.


\section*{Acknowledgements}
This work is supported by the National Key R\&D Program of China (Grant No.
No. 2022YFA1402802). It is also supported in part by the NSFC (Grants No. 11974348 and No. 11834014), and the Strategic Priority
Research Program of CAS (Grants No. XDB28000000, and No. XDB33000000). Z.G.Z. is supported in part by the Training Program of Major Research plan of
the National Natural Science Foundation of China (Grant No. 92165105), and CAS Project for Young Scientists in Basic ResearchGrant No. YSBR-057.



\appendix
\begin{widetext}
\section{Linear spin-wave theory  and magnon Hamiltonian}\label{appa}
Here we rewrite the general two-body spin interaction Hamiltonian,
\begin{equation}\label{hami-heisenberg}
H=\frac{1}{2}\sum_{i,j}^{L}\sum_{n,m}^{N}\sum_{\alpha\beta}{S}^{\alpha}_{i,n}H^{\alpha\beta}_{nm}(i-j){S}^{\beta}_{j,m},
\end{equation}
 By setting a global (reference) coordinates $(\hat{\bm{x}},\hat{\bm{y}},\hat{\bm{z}})$, the local coordinates (spherical coordinates) of each spin relate the global coordinate through
\begin{equation}
\bm{S}_{i,n} = R_{n}(\theta_i,\phi_i)\bm{S}_{0}.
\end{equation}
The classical ground state is identified by treating the quantum mechanical spins operators as classical vectors and minimizing the classical ground-state energy. The magnons are the usual low-energy excitation in ordered magnets, which is considered via the Holstein-Primakoff transformation in local coordinates \cite{petit2011numerical, toth2015linear}
\begin{equation}
\begin{aligned}
&S_{i,n}^{\theta}=\sqrt{\frac{S}{2}}(a_{i,n} + a_{i,n}^\dagger), \quad S_{i,n}^{\phi} = -i\sqrt{\frac{S}{2}}(a_{i,n}-  a_{i,n}^{\dagger}),\\
& S_{i,n}^{r} = S - a_{i,n}^\dagger a_{i,n}.
\end{aligned}
\end{equation}
and we obtain
\begin{equation}
\bm{S}_{i,n}^{\alpha}=\sqrt{\frac{S}{2}}\hat{\bm{u}}_{n} a_{i,n}+\sqrt{\frac{S}{2}}\hat{\bm{u}}_{n}^{ *}a_{i,n}^\dagger + \hat{\bm{z}}_{n}(S - a_{i,n}^\dagger a_{i,n}),
\end{equation}
where $\alpha = x,y,z$, the coefficients $\hat{\bm{u}}_{n}$ and $\hat{\bm{z}}_{n}$ are related to the relative rotation between the global and local coordinates,
that are explicitly written as
\begin{equation}
\begin{aligned}
&\begin{pmatrix}
u_{n}^x \\ u_{n}^y \\ u_{n}^z
\end{pmatrix}=
\begin{pmatrix}
\cos{\theta_n}\cos{\phi_n} + i\sin{\phi_n} \\ \cos{\theta_n}\cos{\phi_n} - i\sin{\phi_n} \\ -\sin{\theta_n}
\end{pmatrix}, \quad \begin{pmatrix}
z_{n}^x \\ z_{n}^y \\z_{n}^z
\end{pmatrix}=
\begin{pmatrix}
\sin{\theta_n}\cos{\phi_n} \\ \sin{\theta_n}\sin{\phi_n} \\ \cos{\theta_n}
\end{pmatrix}.
\end{aligned}
\end{equation}
Expanding the coupling interaction, we obtain
\begin{equation}\label{coup-hop}
\begin{aligned}
S_{i,n}^\alpha S_{j,m}^\beta =& \frac{1}{2} u_{n}^{\alpha *}a_{i,n}^\dagger (u_{m}^\beta a_{j,m} + u_{m}^{\beta *} a_{j,m}^\dagger)+\frac{1}{2} u_{n}^{\alpha }a_{i,n} (u_{m}^{\beta *} a_{j,m}^\dagger + u_{m}^{\beta } a_{j,m}) \\
-& z_{n}^\alpha z_{m}^\beta (a_{i,n}^\dagger a_{i,n} + a_{j,m}^\dagger a_{j,m} ).
\end{aligned}
\end{equation}
By inserting \Eq{coup-hop} into \Eq{hami-heisenberg},  we have
\begin{equation}
\begin{aligned}
H=&\sum_{i,j,n,m} A_{nm}(i-j)a_{i,n}^\dagger a_{j,m}+\frac{1}{2}\sum_{i,j,n,m}\left[B_{nm}(i-j)a_{i,n}^\dagger a_{j,m}^\dagger +\text{h.c.}\right] +2\sum_{i,n,m}C_{nm}a_{i,n}^\dagger a_{i,n}.
\end{aligned}
\end{equation}
In which
\begin{equation}
\begin{aligned}
&A_{nm}(i-j) = \frac{\sqrt{S_n S_m}}{2}\sum_{\alpha\beta}u_{n}^{\alpha *}H_{nm}^{\alpha\beta}(i-j)u_{m}^\beta,\\
&B_{nm}(i-j) = \frac{\sqrt{S_n S_m}}{2}\sum_{\alpha\beta}u_{n}^{\alpha *}H_{nm}^{\alpha\beta}(i-j)u_{m}^{\beta *},\\
&C_{nm} = \delta_{nm}S_l \sum_{\alpha\beta}z_{n}^\alpha \sum_{j}\sum_{l}H_{nm}^{\alpha\beta}(i-j)z_{l}^\beta .
\end{aligned}
\end{equation}
By transforming \Eq{coup-hop} to the reciprocal space, there is
\begin{equation}\label{fourier}
a_{i,n}=(1/\sqrt{L})\sum_{\bm{k}}\exp{[i\bm{k}\cdot (\bm{r}_{i} +\bm{t}_n) ] }a_{\bm{k},n},
\end{equation}
 where $\bm{r}_{i}$ is the position of the $i$th unit cell and $\bm{t}_{m}$ is the relative vector of the $m$th sublattice.
We have $H=\frac{1}{2}\sum_{\bm{k}}\Psi_{\bm{k}}^{\dagger}\mathcal{H}(\bm{k})\Psi_{\bm{k}}$, where
\begin{equation}\label{hami-ap}
\mathcal{H}(\bm{k})=\begin{pmatrix}
A(\bm{k}) - C & B(\bm{k}) \\ B^{\dagger}(\bm{k}) & A^{T}(-\bm{k}) - C
\end{pmatrix}
\end{equation}
 is a 2N$\times$2N bosonic Bogoliubov-de Gennes (BdG) Hamiltonian with the vector boson operator
where $\Psi_{\bm{k}}^{\dagger} = (a_{{\bm{k}},1}^{\dagger}, \cdots , a_{{\bm{k}},N}^{\dagger}, a_{{-\bm{k}},1}, \cdots,  a_{{-\bm{k}},N})$, and the $N\times N$ block matrix is given by
\begin{equation}\label{hami-elem}
\begin{aligned}
&A_{mn}(\bm{k})= \sum_{\alpha,\beta}u_{n}^{\alpha*}\left[\sum_{\bm{d}}e^{-i\bm{k}\cdot\bm{d}}H_{nm}^{\alpha\beta}(\bm{d})  \right]u_{m}^\beta,\\
&B_{mn}(\bm{k})=  \sum_{\alpha,\beta}u_{n}^{\alpha*}\left[\sum_{\bm{d}}e^{-i\bm{k}\cdot\bm{d}}H_{nm}^{\alpha\beta}(\bm{d})  \right]u_{m}^{\beta *},\\
&C_{mn}= \delta_{mn} \sum_{\alpha,\beta}\sum_{l}z_{n}^{\alpha*}\left[\sum_{\bm{d}} H_{nl}^{\alpha\beta}(\bm{d})  \right]z_{l}^\beta.\\
\end{aligned}
\end{equation}
Where $\bm{d}=(\bm{r}_i +\bm{t}_n) -(\bm{r}_j-\bm{t}_m)$ is the difference vector between the $m$th and the $n$th spin. In deriving \Eq{hami-elem}  the relation
\begin{equation}
\sum_{\bm{k}}\sum_{n,m}A_{nm}(\bm{k})a_{\bm{k},n}^\dagger a_{\bm{k},m} = \sum_{\bm{k}}\sum_{n,m}A_{nm}^T(-\bm{k})a_{-\bm{k},n} a_{-\bm{k},m}^\dagger
\end{equation}
is used.
 For collinear ferromagnets, the Hamiltonian \Eq{hami-ap} is block diagonal with identical block which can be reduced to $H=\sum_{\bm{k}}\Psi_{\bm{k}} \mathcal{H}_{\bm{k}} \Psi_{\bm{k}}$ with $\mathcal{H}_{\bm{k}} = (A_{\bm{k}} - C)$ and $\Psi_{\bm{k}}^{\dagger} = (a_{\bm{k},1}^{\dagger},\cdots, a_{\bm{k},N}^{\dagger})$.

\section{Perturbation of electric field in form of minimal coupling}\label{appb}

Now we consider the AC effect.
In \Eq{coup-hop} the first two terms describe the magnon hopping. In the presence of electric field, the magnon acquires a phase while travelling between the $m$th and the $n$th spin on the $i$th and $j$th site, which is given by
\begin{equation}\label{theta-ij}
\theta_{in,jm}=-\frac{g\mu_{B}}{\hbar c^2}\int_{\bm{r}_{i,n}}^{\bm{r}_{j,m}}(\bm{E}(t)\times \hat{\bm{e}}_n)\cdot d\bm{r},
\end{equation}
Noting that \Eq{theta-ij} is general despite the local coordinates dependency on each site. To see this,  let us recall the original expression of AC phase $\theta_{ij}^{\text{AC}}=\frac{1}{\hbar c^2}\int_{\bm{r}_i}^{\bm{r}_j}(\bm{E}(t)\times \bm{\mu})\cdot d\bm{r}$, with $\bm{\mu}$ being the magnetic moment of the magnon. Supposing that the direction and magnitude of the magnon magnetic moment does not change during the hopping processes(i.e., we  neglect the magnon scattering  that involves the angular momentum transfer), then $\theta_{ij}^{\text{AC}}$ is determined by the distance between the sites once the magnetic ground state is fixed, irrespective of the local coordinates. Similar demonstrations can be found in  Ref. \cite{PhysRevB.100.014421} and Ref. \cite{PhysRevB.96.224414}. 
The coupling interaction is modified as
\begin{equation}\label{coup-hop2}
\begin{aligned}
S_{i,n}^\alpha S_{j,m}^\beta =& \frac{1}{2} U_{n}^{\alpha *}a_{i,n}^\dagger (U_{m}^\beta a_{j,m} + U_{m}^{\beta *} a_{j,m}^\dagger)e^{i\theta_{in,jm}}+\frac{1}{2} U_{n}^{\alpha }a_{i,n} (U_{m}^{\beta *} a_{j,m}^\dagger + U_{m}^{\beta } a_{j,m})e^{-i\theta_{in,jm}}\\
&-V_{n}^\alpha V_{m}^\beta (a_{i,n}^\dagger a_{i,n} + a_{j,m}^\dagger a_{j,m} ).
\end{aligned}
\end{equation}
 Supposing that the scale of the spacial variance of $\bm{E}$ is much larger than lattice constant and introducing the effective vector potential $\bm{A}^{E}_n = \frac{1}{c}\bm{E}\times \hat{\bm{e}}_n$, one obtains
\begin{equation}
\begin{aligned}
 \theta_{in,jm}=&\frac{g\mu_B}{\hbar c}\bm{A}^{E}_n\cdot \bm{d}, \\
\end{aligned}
\end{equation}
Following the same procedure, we have
\begin{equation}
\begin{aligned}
A_{nm}(\bm{k})\rightarrow A_{nm}(\bm{k} - \frac{g\mu_B}{\hbar c}\bm{A}^{E}_n ),\\
B_{nm}(\bm{k})\rightarrow B_{nm}(\bm{k} - \frac{g\mu_B}{\hbar c}\bm{A}^{E}_n ).\\
\end{aligned}
\end{equation}
While the matrix $C_{nm}$ is unchanged since for which the magnon hopping is not involved.

Now we consider two special cases: the collinear ferromagnet and collinear antiferromagnet.

\subsection{Collinear ferromagnet}

For ferromagnet due to the absence of the magnon paring $a^{\dagger}_{i,n}a^{\dagger}_{j,m}$ and $a_{i,n}a_{j,m}$, the 2N dimensional basis is reduce to N dimensional $\Psi_{\bm{k}}^\dagger = (a_{\bm{k},1}^{\dagger},\cdots a_{\bm{k},N}^{\dagger})$. Accordingly, the Hamiltonian \Eq{hami-ap} is reduced to $\mathcal{H}_{\bm{k}} = A(\bm{k}) - C$. It is obvious that the global coordinates is same to the local coordinate for each spin, which can be chosen as  that where $z$-direction is identical to the magnetization direction. With this assignment, the magnetic moment of magnon is $-\hbar\hat{\bm{e}}_z$, and the effective vector potential is $\bm{A}^{E}_n = \bm{A}^{E} =\frac{1}{c}\bm{E}\times \hat{\bm{e}}_z$. Consequently, the kernal Hamiltonian becomes $\mathcal{H}(\bm{k}) = A(\bm{k} -\frac{g\mu_B}{\hbar c}\bm{A}^{E})-C$.

\subsection{Collinear antiferromagnet}

To be concrete, we consider a collinear antiferromagnet honeycomb lattice that has two spins in a unit cell, with the Hamiltonian given by
\begin{equation}
H=J_1\sum_{\langle i,j \rangle}\bm{S}_i\cdot \bm{S}_j + D\sum_{\langle\langle ij\rangle\rangle}\xi_{ij}\hat{\bm{z}}\cdot \bm{S}_{i}\times \bm{S}_j +K\sum_i S_{iz}^2.
\end{equation}
In which $J_1 >0$ is the antiferromagnetic exchange interaction, $D$ is the DMI interaction along the $z$ direction, and $\xi_{ij}=1(-1)$ when $\bm{S}_i$ and $\bm{S}_j$ are arranged in a counterclockwise (clockwise) manner. $K<0$ is the easy axis anisotropy. One can choose the local coordinates of 1st spin of the unit cell as the global coordinates, while the local coordinates of the 2nd spin is obtained by a $pi$ rotation about the $x$-axis or $y$-axis of the global coordinates. subsequently, the HP transformation is preformed as
\begin{equation}
\begin{aligned}
\bm{S}_{i,1}^{\alpha}=\sqrt{\frac{S}{2}}\hat{\bm{u}}_{1} a_{i,1}+\sqrt{\frac{S}{2}}\hat{\bm{u}}_{1}^{ *}a_{i,1}^\dagger + \hat{\bm{z}}_{1}(S - a_{i,1}^\dagger a_{i,1}),\\
\bm{S}_{i,2}^{\alpha}=\sqrt{\frac{S}{2}}\hat{\bm{u}}_{2} a_{i,2}+\sqrt{\frac{S}{2}}\hat{\bm{u}}_{2}^{ *}a_{i,2}^\dagger + \hat{\bm{z}}_{2}(S - a_{i,2}^\dagger a_{i,2}),
\end{aligned}
\end{equation}
In which
\begin{equation}
\begin{aligned}
&\begin{pmatrix}
u_{1}^x \\ u_{1}^y \\ u_{1}^z
\end{pmatrix}=
\begin{pmatrix}
1 \\ -i \\ 0
\end{pmatrix},\quad
\begin{pmatrix}
z_{1}^x \\ z_{1}^y \\ z_{1}^z
\end{pmatrix}=
\begin{pmatrix}
0 \\ 0 \\ 1
\end{pmatrix},\\
&\begin{pmatrix}
u_{2}^x \\ u_{2}^y \\ u_{2}^z
\end{pmatrix}=
\begin{pmatrix}
1 \\ i \\ 0
\end{pmatrix},\quad
\begin{pmatrix}
z_{2}^x \\ z_{2}^y \\ z_{2}^z
\end{pmatrix}=
\begin{pmatrix}
0 \\ 0 \\ -1
\end{pmatrix}.
\end{aligned}
\end{equation}
Therefore the magnon Hamiltonian is $H=H^J + H^D + H^K$ with
\begin{equation}
\begin{aligned}
H^{J} =& J_1 S \sum_{\langle i,j\rangle}(a_{i,1}a_{j,2}+a_{i,1}^{\dagger}a_{j,2}^{\dagger}),\\
H^{D} =&-\sum_{\langle\langle i,j\rangle\rangle}iD_2 S(a_{i,1}^{\dagger}a_{j,1} - a_{j,1}^{\dagger}a_{i,1} -a_{i,2}^{\dagger}a_{j,2} + a_{j,2}^{\dagger}a_{i,2}),\\
H^{K} =&\sum_i (3J_1 - K)S (a_{i,1}^{\dagger}a_{i,1} + a_{i,2}^{\dagger}a_{i,2}).
\end{aligned}
\end{equation}
The operator $a_{i,1}$ ($a_{i,1}^\dagger$) annihilates (creates) a magnon with magnetic moment $-\hbar \hat{\bm{e}}_z$, and  the effective vector potential is given as $\bm{A}_{n}^{E} =\bm{A}^{E}= \frac{1}{c}\bm{E}\times \bm{e}_z$. The magnon Hamiltonian in the presence of the electric field is written as
\begin{equation}
\begin{aligned}
H^{J} =& J_1 S \sum_{i,\delta}(a_{i,1}a_{i +\delta,2}e^{i\theta_{r_i,r_i+\delta}^{N}}+\text{h.c.}),\\
H^{D} =&-\sum_{i,\vartheta}iD_2 S(a_{i,1}^{\dagger}a_{i +\vartheta,1}e^{-i\theta_{r_i,r_i+\vartheta}^{NN}} -a_{i,2}^{\dagger}a_{i + \vartheta,2}e^{i\theta_{r_i,r_i+\vartheta}^{NN}} + \text{h.c.}),
\end{aligned}
\end{equation}
where  the phase accumulated along the nearest (second-nearest) neighbor hopping is given as
\begin{equation}
\begin{aligned}
\theta_{r_i,r_i+\delta}^{N}=&-\frac{g\mu_B}{\hbar c^2}\int_{r_i}^{r_i+\delta}(\bm{E}(t)\times \bm{e}_z)\cdot d\bm{r},\\
\theta_{r_i,r_i+\vartheta}^{NN}=&-\frac{g\mu_B}{\hbar c^2}\int_{r_i}^{r_i+\vartheta}(\bm{E}(t)\times \bm{e}_z)\cdot d\bm{r}.
\end{aligned}
\end{equation}   Supposing that the scale of the spacial variance of $\bm{E}$ is much larger than lattice constant one obtains
\begin{equation}
\begin{aligned}
 \theta_{r_i,r_i+\delta}^{N}=&\frac{g\mu_B}{\hbar c}\bm{A}^E\cdot \bm{\delta}, \\
 \theta_{r_i,r_i+\vartheta}^{NN}=&\frac{g\mu_B}{\hbar c}\bm{A}^E\cdot \bm{\vartheta}.
\end{aligned}
\end{equation}
Making use of  the Fourier transformation \Eq{fourier}, one directly obtains $H=\sum_{\bm{k}}\Psi_{\bm{k}}^{\dagger}\mathcal{H}(\bm{k}-\frac{g\mu_B}{\hbar c}\bm{A})\Psi_{\bm{k}}$,
with the Nambu basis given by $\Psi_{\bm{k}}^\dagger = (a_{\bm{k},1}^\dagger,a_{\bm{k},2}^\dagger,a_{-\bm{k},1},a_{-\bm{k},2})$.
 In which $A_{mn}(\bm{k})- C_{mn}$ is diagonal with $A_{11(22)}(\bm{k})- C_{11(22)}= \frac{S}{2}\left[ 3J_1 - K(2S-1)/S \pm D\sum_{\bm{\delta}}2\sin(\bm{k}\cdot \bm{\delta})\right]$, and $B_{mn}(\bm{k})$ is non-diagonal with $B_{12}(\bm{k})=B_{21}^{*}(\bm{k}) = \frac{S}{2}\left[\sum_{\bm{\vartheta}}\exp{(i\bm{k}\cdot \bm{\vartheta})}\right]$.

\section{Gauge transformation and dipole interaction}\label{appc}
In this section we give a gauge transformation to derive perturbed Hamiltonian in form of a dipole interaction.
The single-particle Hamiltonian and the Bloch Hamiltonian satisfy
\begin{equation}
\mathcal{H}_0(-i\bm{\nabla},\bm{r})=e^{i\bm{k}\cdot \bm{r}}\mathcal{H}_0(\bm{k})e^{-i\bm{k}\cdot \bm{r}}.
\end{equation}
According to Sec. II, the perturbed Hamiltonian is written as
\begin{equation}
\mathcal{H}_A = \mathcal{H}_0\left(-i\bm{\nabla} + \frac{g\mu_B}{\hbar c}\bm{A},\bm{r}\right).
\end{equation}
The time-dependent Sch\"{o}rdinger equation is
\begin{equation}
i\hbar\frac{\partial \ket{\psi(\bm{r},t)}}{\partial t}=\mathcal{H}_A\ket{\psi(\bm{r},t)}.
\end{equation}
There is freedom of choice of the phase of the wave functions.
A unitary gauge transformation of $\ket{\psi(\bm{r},t)}$ takes the form
\begin{equation}
\ket{\psi^\prime(\bm{r},t)}=\mathcal{U}(t)\ket{\psi(\bm{r},t)},
\end{equation}
The time-dependent Sch\"{o}rdinger equation transforms as
\begin{equation}\label{sh-e}
i\hbar\frac{\partial \ket{\psi^\prime(\bm{r},t)}}{\partial t}=\left[ \mathcal{U} \mathcal{H}_A \mathcal{U}^\dagger \ket{\psi(\bm{r},t)}+i\hbar\frac{\partial \mathcal{U}(t)}{\partial t} \mathcal{U}^\dagger(t)\right]\ket{\psi^\prime (\bm{r},t)}.
\end{equation}
If the unitary transformation is chosen as $\mathcal{U}(t)= e^{i\mathcal{S}(t)}$ with
\begin{equation}
\mathcal{S}(t)=\frac{g\mu_B}{\hbar c}\bm{A}(t)\cdot \bm{r}.
\end{equation}
For the first term in the r.h.s. of \Eq{sh-e}, by use of the Baker-Campbell-Hausdorff identity
\begin{equation}
\begin{aligned}
e^{i\mathcal{S}}\mathcal{H}_A e^{-i\mathcal{S}} =& \mathcal{H} + i[\mathcal{S},\mathcal{H}_A] -\frac{1}{2}[\mathcal{S},[\mathcal{S},\mathcal{H}_A]] ... + \frac{i^n}{n!}[\mathcal{S},  ...,  [\mathcal{S},\mathcal{H}_A]] + ... ,
\end{aligned}
\end{equation}
it is obtained that
 \begin{equation}
\mathcal{U}(t)\mathcal{H}_A \mathcal{U}^\dagger(t)=\mathcal{H}_0.
\end{equation}
For the second term in the r.h.s. of \Eq{sh-e}, by introducing the effective electric field $\tilde{\bm{E}} = -\partial_t \bm{A}$, it is straightforward to show that
\begin{equation}
i\hbar \frac{\partial \mathcal{U}(t)}{\partial t}\mathcal{U}^\dagger(t)=\frac{g\mu_B}{c}\tilde{\bm{E}}(t)\cdot \bm{r}.
\end{equation}
Then the Hamiltonian in velocity gauge transform to
\begin{equation}
\mathcal{H}_{E}(t)=\mathcal{H}_0\left(\bm{k}\right)+\frac{g\mu_B}{c}\tilde{\bm{E}}(t)\cdot \bm{r}.
\end{equation}

\section{Magnon spin current	}\label{appd}
Now we define the magnon spin current. Because the $z$ component of the total spin is conserved, the local  magnon spin density (LMSD)  is $n_z (\bm{r}_i)=\hbar\sum_{m}\bm{z}_m a_{i,m}^{\dagger}a_{i,m}$ satisfying the continuity equation. The Fourier transformation of the LMSD is written as
\begin{equation}
n_z (\bm{r}_i)=\frac{\hbar}{N}\sum_{\bm{k}\bm{q}m}\bm{z}_m e^{-i\bm{q}\cdot (\bm{r}_{i} +\bm{t}_{m})}a_{\bm{k}+\bm{q},m}^{\dagger}a_{\bm{k},m}.
\end{equation}
The Heisenberg equation of motion for $a_{\bm{k},m}$ is used to derive the equation of motion for $n_z (\bm{r}_i)$:
\begin{equation}
\begin{aligned}
 \dot{a}_{\bm{k},m} =& \frac{1}{i\hbar}[a_{\bm{k},m}, H]\\
 =&\frac{1}{i\hbar} \sum_{n}( A_{\bm{k},mn} a_{\bm{k},n} +A_{\bm{k},mn} a_{\bm{k},n} +B_{\bm{k},mn}^* a_{-\bm{k},n}), \\
  =&\frac{1}{i\hbar} \sum_{n}( 2A_{\bm{k},mn} a_{\bm{k},n}  +B_{\bm{k},mn} a_{-\bm{k},n}^\dagger +B_{-\bm{k},nm} a_{-\bm{k},n}^\dagger), \\
\end{aligned}
\end{equation}
where $H$ is given by \Eq{hami-ap}, and we obtain:
\begin{equation}
\begin{aligned}
\frac{\partial n_z (\bm{r}_i)}{\partial t}=& \frac{1}{i N}\sum_{\bm{k}\bm{q}mn}e^{-i\bm{q}\cdot \bm{r}_{i,m}} \bm{z}_m \left[-\left( 2A_{\bm{k+q},nm} a_{\bm{k+q},n}^\dagger  +(B^{*}_{\bm{k+q},mn}+B^{*}_{\bm{-k-q},nm}) a_{-\bm{k-q},n}^\dagger \right)a_{\bm{k},m}\right. \\
&\left.  +a_{\bm{k+q},m}^\dagger\left( 2A_{\bm{k},mn} a_{\bm{k},n}  +(B_{\bm{k},mn} + B_{-\bm{k},nm}) a_{-\bm{k},n}^\dagger\right)\right].\\
\end{aligned}
\end{equation}
In the case $\sum \bm{\delta}=0$ and assuming the long-wavelength limit $\bm{q}\rightarrow 0$, the magnon spin current is obtained by use of the continuity equation $\partial n_{z\bm{q}} /\partial t + i\bm{q}\cdot \bm{J}_{s,z} = 0$, which is given as
\begin{equation}\label{js-1}
\begin{aligned}
\bm{J}_{s,z} =& \hbar\sum_{\bm{k}mn}\bm{z}_m\left( \frac{\partial A_{\bm{k},mn}}{\partial \bm{k}}a_{\bm{k},m}^{\dagger}a_{\bm{k},n}+\frac{\partial A_{-\bm{k},nm}}{\partial \bm{k}}a_{-\bm{k},m}a_{-\bm{k},n}^{\dagger} \right.\\
+ & \left.\frac{\partial B_{\bm{k},mn}}{\partial \bm{k}}a_{\bm{k},m}^\dagger a_{-\bm{k},n}^\dagger +\frac{\partial B_{\bm{k},nm}^* }{\partial \bm{k}}a_{-\bm{k},m} a_{\bm{k},n}\right)\\
=&\hbar\sum_{\bm{k}}\Psi_{\bm{k}}^\dagger \mathcal{Z}\frac{ \partial \mathcal{H}_{\bm{k}}}{\partial \bm{k}}\Psi_{\bm{k}}.
\end{aligned}
\end{equation}
Noting that when long-wavelength limit is invoked, the continuous translational symmetry is respected by removing the dependences
on the lattice structure long-wavelength limit. In the low
temperature we considered, the spin-wave excitation is dominated by the long-wave modes, yielding the justification of the long-wave approximation.
In \Eq{js-1} we define the diagonal matrix $\mathcal{Z}$
\begin{equation}\label{direc-ma}
\mathcal{Z}=\text{diag}(z_1,\cdots,z_N, z_1,\cdots,z_N).
\end{equation}
Making use of \Eq{trans}, we have
\begin{equation}\label{js}
\begin{aligned}
\bm{J}_{s,z} =\hbar\sum_{\bm{k}}\Phi_{\bm{k}}^\dagger \Sigma_z \frac{ \widetilde{\partial \mathcal{Z}\mathcal{H}_{\bm{k}}}}{\partial \bm{k}}\Phi_{\bm{k}}.
\end{aligned}
\end{equation}
It is found that
\begin{equation}
\begin{aligned}
\frac{ \widetilde{\partial \mathcal{Z}\mathcal{H}_{\bm{k}}}}{\partial \bm{k}}=\frac{\partial \widetilde{\mathcal{Z}\mathcal{H}}_{\bm{k}}}{\partial \bm{k}} - \widetilde{\mathcal{Z}\mathcal{H}}_{\bm{k}}U_{\bm{k}}^{-1}\frac{\partial U_{\bm{k}}}{\partial \bm{k}} - \frac{\partial U_{\bm{k}}^{-1}}{\partial \bm{k}}U_{\bm{k}}\widetilde{\mathcal{Z}\mathcal{H}}_{\bm{k}},
\end{aligned}
\end{equation}
and we obtain
\begin{equation}\label{js2}
\begin{aligned}
\bm{J}_{s,z} =\hbar\sum_{\bm{k}}\Phi_{\bm{k}}^\dagger \Sigma_z \left( \frac{\partial \widetilde{ \mathcal{Z}\mathcal{H}_{\bm{k}}}}{\partial \bm{k}}-i[\bm{\mathcal{A}}_{\bm{k}},\widetilde{ \mathcal{Z}\mathcal{H}_{\bm{k}}}]\right)\Phi_{\bm{k}},
\end{aligned}
\end{equation}
where the Berry connection is defined by
 \begin{equation}\label{berry-cone}
 \bm{\mathcal{A}}_{\bm{k}} = iU_{\bm{k}}^{-1}\frac{\partial U_{\bm{k}}}{\partial \bm{k}} = i\Sigma_z U_{\bm{k}}^{\dagger}\Sigma_z \frac{\partial U_{\bm{k}}}{\partial \bm{k}}.
\end{equation}
 Noting that the Berry connection given in \Eq{berry-cone} is different from $iU_{\bm{k}}^{\dagger}\frac{\partial U_{\bm{k}}}{\partial \bm{k}}$. This is because that the Bogoliubov transformation is generally not unitary \cite{xiao2009theory}.

 For collinear ferromagnets, owing to the reduction of 2N dimensional basis to the N dimensional the basis $\Psi_{\bm{k}}^{\dagger} = (a_{\bm{k},1}^{\dagger},\cdots, a_{\bm{k},N}^{\dagger})$, the direction matrix $\mathcal{Z}$ in \Eq{direc-ma} becomes a unit matrix and the magnon spin current operator in \Eq{js2} becomes
\begin{equation}
\begin{aligned}
\bm{J}_s =\hbar\sum_{\bm{k}}\Phi_{\bm{k}}^\dagger \left( \frac{\partial { \mathcal{H}_{\bm{k}}}}{\partial \bm{k}}-i[\bm{\mathcal{A}}_{\bm{k}},{ \mathcal{H}_{\bm{k}}}]\right)\Phi_{\bm{k}}.
\end{aligned}
\end{equation}

\section{Magnon spin photocurrent in collinear ferromagnets}\label{appe}

In this section we derive the equations of motion for the magnon density matrix in the presence of the AC phase induced by the electric field of light. A similar method for electron second-order optical responses is used to calculate the electric bulk photogalvanic effect \cite{PhysRevX.11.011001, PhysRevX.10.041041}. We give the formalism of nonlinear responses for the magnon spin photocurrent by standard perturbation technique. According to Eq. (7) in the maintext, the recursion equation can be rewritten as
\begin{equation}
\begin{aligned}
\rho_{mn}^{(n+1)}(\omega)=&\frac{g\mu_{B}}{\hbar c^2}d_{mn}(\omega)\int \frac{d\omega_1}{2\pi} \omega_1   \epsilon^{\alpha_1 z \beta} E^{\alpha_{1}} D^{\beta} [ \rho^{(n)}(\omega-\omega_1)]_{mn},
\end{aligned}
\end{equation}
where $\epsilon^{\alpha_1 z \beta}$ the Levi-Civita symbol, and summation is implied over repeated spatial indices.
In deriving the second-order reduced density matrix, it can be divided into terms originating from the intraband (i) and interband (e) components of $\mathcal{D}_{\text{opt}}$ operator:
\begin{equation}\label{rdm2}
\begin{aligned}
\rho^{(2)}_{ mn }(\omega)=& \nu_c\sum_{X} \int \frac{d\omega_1 d\omega_2}{(2\pi)^2}  {\omega_1 \omega_2} E^{\alpha_1}(\omega_1)E^{\alpha_2}(\omega_2) \epsilon^{\alpha_1 z \beta} \epsilon^{\alpha_2 z \gamma} \varrho^{X,\beta\gamma}_{mn}(\omega_1,\omega_2),
\end{aligned}
\end{equation}
with the constant $\frac{g^2 \mu_{B}^2}{\hbar^2 c^4}$ is noted as $\nu_c$.
Then, the contribution of each term in \Eq{rdm2} to the spin photoconductivities are noted as $\sigma_{X}^{\alpha\alpha_1 \alpha_2}$ for $X=$ $\text{ii}$, $\text{ie}$, $\text{ei}$ and $\text{ee}$.
Keeping in mind that the energy conservation $\omega = \omega_1 + \omega_2 $ is always satisfied,
the expressions for each term are obtained as
\begin{equation}\label{rdm-4}
\begin{aligned}
\varrho^{\text{ii},\beta\gamma}_{  mn}(\omega_1,\omega_2)=& -d_{mn}(\omega)d_{mn}(\omega_2) (\partial^{\beta}\partial^{\gamma}g_{  m})\delta_{mn},\\
\varrho^{\text{ei},\beta\gamma}_{  mn}(\omega_1,\omega_2)=& -id_{mn}(\omega)d_{mm}(\omega_2)\mathcal{A}_{  mn}^{\gamma}\partial^{\beta}g_{nm},\\
\varrho^{\text{ie},\beta\gamma}_{  mn}(\omega_1,\omega_2)=& -id_{mn}(\omega)\left(\partial^{\gamma}\left[d_{mn}(\omega_2)  \mathcal{A}^{\beta}_{mn}g_{nm}\right] -i(\mathcal{A}^{\gamma}_{mm}-\mathcal{A}^{\gamma}_{nn})\left[d_{mn}(\omega_2)  \mathcal{A}^{\beta}_{mn}g_{nm}\right]\right),\\
\varrho^{\text{ee},\beta\gamma}_{  mn}(\omega_1,\omega_2)=& \sum_{p\neq m,n} d_{mn}(\omega)\left[\mathcal{A}_{  mp}^{\gamma}d_{p n}(\omega_2)\mathcal{A}_{  pn}^{\beta}g_{np}  -\mathcal{A}_{  pn}^{\gamma}d_{mp}(\omega_2)\mathcal{A}_{  mp}^{\beta}g_{pm}\right].
\end{aligned}
\end{equation}
Where we define  $g_{mn}=g_m - g_n $. Note that $\varrho^{ii}$ and $\varrho^{ei}$ include the derivative to the Bose-Einstein distribution.
 Together with \Eq{js2} and \Eq{rdm-4}, the second-order magnon spin current is given as
\begin{equation}
J^{(2),\alpha}_{s}(\omega_1,\omega_2) = \sum_{X}E^{\beta}(\omega_1)E^{\gamma}(\omega_2)\chi_{X}^{\alpha\beta\gamma}(\omega_1,\omega_2) ,
\end{equation}
with
\begin{align}
\chi^{\alpha\alpha_1 \alpha_2}_{\text{ii}}=&  -\frac{\nu_c}{2} \int [d \bm{k} ]\sum_m \epsilon^{\alpha_1 z \beta}\epsilon^{\alpha_2 z \gamma}{\omega_1\omega_2} J_{s,mm}^{\alpha}d_{mm}(\omega_2) d_{mm}(\omega) \partial^{\beta}\partial^{\gamma}g_{  m} \\
+& [(\alpha_1 , \omega_1)\leftrightarrow (\alpha_2, \omega_2)],\label{chi-drude}\\
\chi^{\alpha\alpha_1 \alpha_2}_{\text{ei}}=& -\frac{\nu_c}{2} \int [d \bm{k} ]\sum_{m,n}i\epsilon^{\alpha_1 z \beta}\epsilon^{\alpha_2 z \gamma}{\omega_1\omega_2}J_{s,mn}^{\alpha} d_{nm}(\omega) d_{mm}(\omega_2)\mathcal{A}_{  nm}^{\gamma}\partial^{\beta}g_{nm} \\
+& [(\alpha_1 , \omega_1)\leftrightarrow (\alpha_2, \omega_2)], \label{bcd}\\
\notag\chi^{\alpha\alpha_1 \alpha_2}_{\text{ie}}=& -\frac{\nu_c}{2}\int [d \bm{k} ]\sum_{m,n}i \epsilon^{\alpha_1 z \beta}\epsilon^{\alpha_2 z \gamma}{\omega_1\omega_2}J_{s,mn}^{\alpha} d_{nm}(\omega) \left(\partial^{\beta}\left[d_{nm}(\omega_2)  \mathcal{A}^{\gamma}_{nm}g_{nm}\right] -i(\mathcal{A}^{\beta}_{nn}-\mathcal{A}^{\beta}_{mm})\right.\\
 &\left. \times\left[d_{nm}(\omega_2)  \mathcal{A}^{\gamma}_{nm}g_{nm}\right]\right)+ [(\alpha_1 , \omega_1)\leftrightarrow (\alpha_2, \omega_2)],\\
\notag\chi^{\alpha\alpha_1 \alpha_2}_{\text{ee}}= & \frac{\nu_c}{2}\int [d \bm{k} ]\sum_{m,n,p}\epsilon^{\alpha_1 z \beta}\epsilon^{\alpha_2 z \gamma}{\omega_1\omega_2}J_{s,mn}^{\alpha} d_{nm}(\omega) \left[d_{pm}(\omega_2)\mathcal{A}_{  np}^{\beta}\mathcal{A}_{  pm}^{\gamma}g_{mp} \right. \\
&-\left.d_{np}(\omega_2)\mathcal{A}_{  pm}^{\beta}\mathcal{A}_{  np}^{\gamma}g_{pn}\right]
+ [(\alpha_1 , \omega_1)\leftrightarrow (\alpha_2, \omega_2)].
\end{align}
Note that the conductivity tensors are symmetrized by the indices and frequencies of electric fields. This is because the physical observables should not be affected by an arbitrary permutation of applied external fields, that is, the conductivity tensor has intrinsic permutation symmetry \cite{PhysRevX.11.011001, tsirkin2022separation}. It is clearly seen that the magnon photocurrent is quite different from the case of charge photocurrent. For the magnon photocurrent the electric field is restricted into the plane perpendicular to the axis of the magnon spin, while for the charge case the direction of the electric field can be arbitrary.

\subsection{Drude spin current}
According to \Eq{chi-drude}, $\chi^{\alpha\alpha_1 \alpha_2}_{\text{ii}}$ only involves the intraband contribution, which is given by
\begin{equation} \label{eta-drude}
\begin{aligned}
\chi^{\alpha\alpha_1 \alpha_2}_{\text{ii}}=& - \lim_{\omega \rightarrow 0}\frac{\nu_c}{2\hbar^2 \omega} \int [d \bm{k} ]\sum_m \left(\epsilon^{\alpha_1 z \beta}\epsilon^{\alpha_2 z \gamma}\frac{1}{-\Omega} +\epsilon^{\alpha_2 z \beta}\epsilon^{\alpha_1 z \gamma}\frac{1}{\omega + \Omega}\right)\Omega (\omega + \Omega) J_{s,mm}^{\alpha}  \partial^{\beta}\partial^{\gamma}g_{  m}.
\end{aligned}
\end{equation}
It describes the drift movement of magnons driving by the optical field, and is noted as the Drude contribution.
Following the definitions
\begin{equation}
\begin{aligned}
&\eta^{\alpha\alpha_1 \alpha_2} =\frac{1}{2}\text{Re}(\chi^{\alpha\alpha_1 \alpha_2} + \chi^{\alpha\alpha_2 \alpha_1}), \\
&\kappa^{\alpha\alpha_1 \alpha_2} = \frac{1}{2}\text{Im}(\chi^{\alpha\alpha_1 \alpha_2}- \chi^{\alpha\alpha_2 \alpha_1}),
\end{aligned}
\end{equation}
we obtain the LP spin photocurrent conductivity
\begin{equation}
\eta^{\alpha\alpha_1 \alpha_2}_{\text{D}}= \frac{\nu_c}{2\hbar^2 } \int [d \bm{k} ]\sum_m  \left(\epsilon^{\alpha_1 z \beta}\epsilon^{\alpha_2 z \gamma} + \epsilon^{\alpha_2 z \beta}\epsilon^{\alpha_1 z \gamma}\right)  v_{m}^{\alpha}\partial^{\beta}\partial^{\gamma}g_{  m}.
\end{equation}
In \Eq{eta-drude} we use the relation $J_{s,mm}^{\alpha} = \hbar v_{m}$ with $v_{m}=\frac{1}{\hbar}\partial^{\alpha}\varepsilon_m$ being the group velocity.
To simplify the formulas, we define the symmetric combination of the Levi-Civita symbols as
\begin{equation}\label{es}
E_{S}^{\alpha_1\alpha_2\beta\gamma}=\epsilon^{\alpha_1 z \beta}\epsilon^{\alpha_2 z \gamma} + \epsilon^{\alpha_2 z \beta}\epsilon^{\alpha_1 z \gamma}
\end{equation}
 and the antisymmetric combination one as
 \begin{equation}\label{ea}
 E_{A}^{\alpha_1\alpha_2\beta\gamma}=\epsilon^{\alpha_1 z \beta}\epsilon^{\alpha_2 z \gamma} - \epsilon^{\alpha_2 z \beta}\epsilon^{\alpha_1 z \gamma}.
\end{equation}
and we have
\begin{equation}\label{eta-drude-1}
\begin{aligned}
\eta^{\alpha\alpha_1 \alpha_2}_{\text{D}}= & \frac{\nu_c}{2\hbar } \int [d \bm{k} ]\sum_m  E_{S}^{\alpha_1\alpha_2\beta\gamma} v_{m}^{\alpha}\partial^{\beta}\partial^{\gamma}g_{  m}.
\end{aligned}
\end{equation}
Note that the integrand $v_{m}^{\alpha}\partial^{\beta}\partial^{\gamma}g_{  m}$ in \Eq{eta-drude-1} is symmetric by exchanging the external field indices $\beta$ and $\gamma$, hence $E_{S}^{\alpha_1\alpha_2\beta\gamma} = 2$ for $\alpha_1 = \alpha_2$ while $E_{S}^{\alpha_1\alpha_2\beta\gamma} = -2$ for $\alpha_1 \neq \alpha_2$, and \Eq{eta-drude-1} can be rewritten as
\begin{equation}\label{eta-drude-2}
\begin{aligned}
\eta^{\alpha\alpha_1 \alpha_2}_{\text{D}} =\varsigma\frac{\nu_c}{\hbar } \int [d \bm{k} ]\sum_m   v_{m}^{\alpha}\partial^{\alpha_1}\partial^{\alpha_2}g_{  m},
\end{aligned}
\end{equation}
where we define $\varsigma = 1$ for  $\alpha_1 = \alpha_2$ and $\varsigma = -1$ for  $\alpha_1 \neq \alpha_2$. Noting that for the LP Drude response it is independent to the frequency of the light.

The CP-component is given as
\begin{equation}\label{kappa-drude}
\begin{aligned}
\kappa^{\alpha\alpha_1 \alpha_2}_{\text{ii}}= &- \lim_{\omega \rightarrow 0}\frac{i\nu_c}{2\hbar \omega} \int [d \bm{k} ]\sum_m E_{S}^{\alpha_1\alpha_2\beta\gamma}(\frac{1}{-\Omega} -\frac{1}{\omega + \Omega})\Omega (\omega + \Omega) v_{m}^{\alpha}  \partial^{\beta}\partial^{\gamma}g_{  m}\\
= &- \lim_{\omega \rightarrow 0}\frac{i\nu_c}{\hbar \omega} \int [d \bm{k} ]\sum_m E_{S}^{\alpha_1\alpha_2\beta\gamma}\Omega v_{m}^{\alpha}  \partial^{\beta}\partial^{\gamma}g_{  m}.
\end{aligned}
\end{equation}
For the CP responses the external field indices are restricted with $\alpha_1 \neq \alpha_2$, making use of the exchanging symmetric property of the integrand $v_{m}^{\alpha}\partial^{\beta}\partial^{\gamma}g_{  m}$, it is easy to verify  that the CP-response vanishes.

\subsection{Berry curvature dipole (BCD) spin current}
In this subsection, we consider the spin photocurrent originating from $\chi^{\alpha\alpha_1 \alpha_2}_{s,\text{ei}}$. For the dc current, it is given as
\begin{equation}
\begin{aligned}
\chi^{\alpha\alpha_1 \alpha_2}_{\text{ei}}=&\lim_{\omega \rightarrow 0} \frac{\nu_c}{2\hbar} \int  [d \bm{k} ]\sum_{m\neq n}(\epsilon^{\alpha_1 z \beta}\epsilon^{\alpha_2 z \gamma}\frac{1}{-\Omega} +\epsilon^{\alpha_2 z \beta}\epsilon^{\alpha_1 z \gamma}\frac{1}{\omega + \Omega}) \Omega(\omega + \Omega)\mathcal{A}_{mn}^{\alpha} \mathcal{A}_{  nm}^{\gamma}\partial^{\beta}g_{nm},
\end{aligned}
\end{equation}
where we used the relation $\lim_{\omega \rightarrow 0} d_{nm}(\omega) J_{s,mn}^{\mu} =i\hbar^{-1}\mathcal{A}_{mn}^{\mu} $  for $m\neq n$. The LP-photocurrent tensor is
\begin{equation}\label{eta-ei}
\begin{aligned}
\eta^{\alpha\alpha_1 \alpha_2}_{\text{ei}}=& \lim_{\omega \rightarrow 0}\frac{\nu_c\omega}{2\hbar} \int  [d \bm{k} ]\sum_{m\neq n}(\epsilon^{\alpha_1 z \beta}\epsilon^{\alpha_2 z \gamma}+\epsilon^{\alpha_2 z \beta}\epsilon^{\alpha_1 z \gamma}) \mathcal{A}_{mn}^{\alpha} \mathcal{A}_{  nm}^{\gamma}\partial^{\beta}g_{nm}\\
=&\lim_{\omega \rightarrow 0}\frac{\nu_c\omega}{2\hbar} \int  [d \bm{k} ]\sum_{m\neq n}E_{S}^{\alpha_1\alpha_2\beta\gamma} (\mathcal{A}_{mn}^{\alpha} \mathcal{A}_{  nm}^{\gamma}-\mathcal{A}_{mn}^{\gamma} \mathcal{A}_{  nm}^{\alpha})\partial^{\beta}g_{m}.
\end{aligned}
\end{equation}
It can be seen that the LP response vanishes for the dc current.
The CP-photocurrent tensor is derived as
\begin{equation}
\begin{aligned}\label{kappa-ei}
\kappa^{\alpha\alpha_1 \alpha_2}_{\text{BCD}}=&i\frac{\nu_c \Omega}{2\hbar} \int  [d \bm{k} ] \sum_{m\neq n}(\epsilon^{\alpha_1 z \beta}\epsilon^{\alpha_2 z \gamma}-\epsilon^{\alpha_2 z \beta}\epsilon^{\alpha_1 z \gamma}) \mathcal{A}_{mn}^{\alpha} \mathcal{A}_{  nm}^{\gamma}\partial^{\beta}g_{nm}\\
=&\frac{\nu_c \Omega}{2\hbar^2} \int  [d \bm{k} ]\sum_{m}E_{A}^{\alpha_1\alpha_2\beta\gamma} i\epsilon^{\alpha\gamma\tau}\partial^{\beta}\Omega_{m}^{\tau}g_{m},
\end{aligned}
\end{equation}
The magnon Berry curvature is defined as
\begin{equation}\label{berry-curv}
\Omega_{m}^{\tau} = \frac{i}{2}\epsilon^{\alpha\gamma\tau}\sum_{n\neq m}(\mathcal{A}_{mn}^{\alpha} \mathcal{A}_{  nm}^{\gamma}-\mathcal{A}_{m,n}^{\gamma} \mathcal{A}_{  nm}^{\alpha})
\end{equation}
By use of \Eq{berry-curv} and partial integration, \Eq{kappa-ei} can be rewritten as
which depends on the frequency of incident light as $O(\Omega)$.
\begin{equation}\label{kappa-bcd}
\begin{aligned}
\kappa^{\alpha\alpha_1 \alpha_2}_{\text{BCD}}=&\frac{\nu_c \Omega}{2\hbar} \int  [d \bm{k} ]\sum_{m}E_{A}^{\alpha_1\alpha_2\beta\gamma} \epsilon^{\alpha\gamma\tau}\partial^{\beta}\Omega_{m}^{\tau}g_{m}\\
=&\frac{\nu_c \Omega}{2\hbar} \int  [d \bm{k} ]\sum_{m} (\epsilon^{\alpha\alpha_2\tau}\partial^{\alpha_1} - \epsilon^{\alpha\alpha_1\tau}\partial^{\alpha_2} )\Omega_{m}^{\tau}g_{m},
\end{aligned}
\end{equation}
It is clearly seen that it comes from the dipole of Berry curvature in momentum space. The BCD contribution is classified as a CP photocurrent. The BCD induced nonlinear current has been extensively studied in electric systems. In  bosonic systems, the magnon BSC driven by temperature gradient has been proposed \cite{PhysRevResearch.4.013186}. Here we reveal the BCD contribution to magnon spin current generate via light method.

\subsection{Injection and rectification spin currents}
Now we study the spin photocurrent comes from $\chi_{\text{ee}}$. First we consider the contribution from the diagonal component of $\chi_{\text{ee}}$ with $m=n$, denoted as $\chi_{\text{ee,d}}$, which is given by
\begin{eqnarray}
\chi^{\alpha\alpha_1 \alpha_2}_{\text{ee,d}} &= & \lim_{\omega \rightarrow 0}\frac{\nu_c \hbar}{2 \omega}\int [d \bm{k} ]\sum_{m \neq p}\epsilon^{\alpha_1 z \beta}\epsilon^{\alpha_2 z \gamma}{\omega_1\omega_2}\Delta_{mp}^{\alpha} d_{p m}(\omega_2)\mathcal{A}_{  mp}^{\beta} \mathcal{A}_{  pm}^{\gamma}g_{mp}+ [(\alpha_1 , \omega_1)\leftrightarrow (\alpha_2, \omega_2)] \notag\\
&=& \lim_{\omega \rightarrow 0} \frac{\nu_c\hbar}{2\hbar \omega}\int [d \bm{k} ]\sum_{m \neq p}\left(\epsilon^{\alpha_1 z \beta}\epsilon^{\alpha_2 z \gamma}\frac{1}{-\hbar\Omega -\varepsilon_{pm}} + \epsilon^{\alpha_2 z \beta}\epsilon^{\alpha_1 z \gamma}\frac{1}{\hbar\omega + \hbar\Omega - \varepsilon_{pm}}\right) \notag\\
& \times & \Omega(\omega + \Omega)\Delta_{mp}^{\alpha} \mathcal{A}_{  mp}^{\beta} \mathcal{A}_{  pm}^{\gamma}g_{mp}.
\label{chi-alpha12}
\end{eqnarray}
where $\Delta_{mp}^{\alpha}$ is the interband-transition of the velocity. Making use of \Eq{es} and \Eq{ea}, we have
\begin{eqnarray}
\eta^{\alpha\alpha_1 \alpha_2}_{\text{ee,d}}&=&\lim_{\omega \rightarrow 0} \frac{\nu_c\hbar}{2 \omega}\int [d \bm{k} ]\sum_{m \neq p}E_{S}^{\alpha_1\alpha_2\beta\gamma}\left(\frac{1}{-\hbar\Omega -\varepsilon_{pm}} + \frac{1}{\hbar\omega + \hbar\Omega - \varepsilon_{pm}}\right) \notag\\
& \times & \Omega(\omega + \Omega)\Delta_{mp}^{\alpha} \mathcal{A}_{  mp}^{\beta} \mathcal{A}_{  pm}^{\gamma}g_{mp} \notag\\
&=& \lim_{\omega \rightarrow 0} \frac{\nu_c\hbar}{2 \omega}\int [d \bm{k} ]\sum_{m \neq p}E_{S}^{\alpha_1\alpha_2\beta\gamma}\left(\frac{1}{-\hbar\Omega -\varepsilon_{pm}} \mathcal{A}_{  mp}^{\beta} \mathcal{A}_{  pm}^{\gamma}+ \frac{1}{\hbar\omega + \hbar\Omega + \varepsilon_{pm}} \mathcal{A}_{  pm}^{\beta} \mathcal{A}_{  mp}^{\gamma}\right) \notag\\
& \times & \Omega(\omega + \Omega)\Delta_{mp}^{\alpha} g_{mp}.
\label{eta-eed-o}
\end{eqnarray}
Because of the prefactor $1/\omega$, \Eq{eta-eed-o} diverges with the limit $\omega \rightarrow 0$. Therefore $O(\omega)$ and $O(\omega^0)$ are retained. Perform the Taylor expansion
\begin{equation}\label{taylor}
\frac{1}{\hbar\omega + \Omega + \varepsilon_{nm}} = \frac{1}{\hbar\Omega + \varepsilon_{nm}} - \frac{\omega}{(\hbar\Omega + \varepsilon_{nm})^2}+ O(\omega^2),
\end{equation}
and by use of the Cauchy principal integral
\begin{equation}\label{cauchy}
\frac{1}{\hbar\Omega + i0^{+} + \varepsilon_{nm}} = \text{P}(\frac{1}{\hbar\Omega - \varepsilon_{mn}}) - i\pi\delta(\hbar\Omega - \varepsilon_{mn}),
\end{equation}
where $P$ denotes the principal value. We have
\begin{eqnarray}
\eta^{\alpha\alpha_1 \alpha_2}_{\text{ee,d}} &=&-\lim_{\omega \rightarrow 0} \frac{i\pi\nu_c\hbar}{2 \omega}\int [d \bm{k} ]\sum_{m \neq n}E_{S}^{\alpha_1\alpha_2\beta\gamma}\delta(\hbar\Omega - \varepsilon_{mn}) \Omega(\omega + \Omega)  \notag\\
&\times & \Delta_{mn}^{\alpha} (\mathcal{A}_{  mn}^{\beta} \mathcal{A}_{  nm}^{\gamma} +  \mathcal{A}_{  nm}^{\beta} \mathcal{A}_{  mn}^{\gamma})g_{mn} \notag\\
&-& \lim_{\omega \rightarrow 0} \frac{\nu_c\hbar}{2 \omega}\int [d \bm{k} ]\sum_{m \neq n}E_{S}^{\alpha_1\alpha_2\beta\gamma}\frac{\omega}{(\hbar\Omega - \varepsilon_{mn})^2} \Omega(\omega + \Omega)\Delta_{mn}^{\alpha} \mathcal{A}_{  nm}^{\beta} \mathcal{A}_{  mn}^{\gamma} g_{mn}.
\label{eta-ee-d}
\end{eqnarray}
Define the quantum metric and the Berry curvature
\begin{align}
G_{mn}^{\beta\gamma} &=\frac{1}{2}( \mathcal{A}_{  mn}^{\beta} \mathcal{A}_{  nm}^{\gamma } + \mathcal{A}_{  mn}^{\gamma} \mathcal{A}_{  nm}^{\beta}),\\
\Omega_{mn}^{\beta\gamma} &=i( \mathcal{A}_{  mn}^{\beta} \mathcal{A}_{  nm}^{\gamma } - \mathcal{A}_{  mn}^{\gamma} \mathcal{A}_{  nm}^{\beta}).
\end{align}
The first term in \Eq{eta-ee-d} diverges as $\omega \rightarrow 0$, which is recognized as the injection current.  Considering that the band-resolved quantum metric is symmetric with the indices permutation $\beta \leftrightarrow \gamma$, the injection current is
\begin{equation}\label{eta-inj}
\begin{aligned}
\eta^{\alpha\alpha_1 \alpha_2}_{\text{Inj}}=&-\lim_{\omega \rightarrow 0} \frac{i\pi\nu_c\hbar}{ \omega}\int [d \bm{k} ]\sum_{m \neq n}E_{S}^{\alpha_1\alpha_2\beta\gamma}\delta(\hbar\Omega - \varepsilon_{mn}) \Omega^2 \Delta_{mn}^{\alpha} G_{mn}^{\beta\gamma}g_{mn}\\
=&-\lim_{\omega \rightarrow 0}\varsigma \frac{i2\pi\nu_c}{ \omega}\int [d \bm{k} ]\sum_{m \neq n}\delta(\hbar\Omega - \varepsilon_{mn}) \Omega^2 \Delta_{mn}^{\alpha} G_{mn}^{\alpha_1\alpha_2}g_{mn}.
\end{aligned}
\end{equation}
Noting that the LP injection current diverges in the dc limit. This seemingly unphysical behavior can be eliminated by the introducing the phenomenological scattering rate $\Gamma$, and the calculations are carried out by the shift of the poles, i.e., the matrix $d_{mn}(\omega)$ is modified as
\begin{equation}\label{dmatrix}
d_{mn}(\omega) = \frac{1}{\hbar\omega - \varepsilon_{mn}+ i\varsigma}.
\end{equation}
In \Eq{dmatrix} the relaxation time approximation is invoked with $\varsigma = \alpha\omega =1/\tau$, $\alpha$ is the Gilbert damping constant and $\tau$ is the relaxation time.
The delta function holds the property
\begin{equation}\label{delta}
\delta(\hbar\Omega - \varepsilon_{mn})=\lim_{\varsigma\rightarrow 0}\frac{1}{\pi}\frac{\varsigma}{(\hbar\Omega -\varepsilon_{mn})^2 +\varsigma^2}.
\end{equation}
By use of \Eq{delta}, the injection current is given by
\begin{equation}
\begin{aligned}
\eta^{\alpha\alpha_1 \alpha_2}_{\text{Inj}}
=&-\varsigma {2\nu_c\hbar}\int [d \bm{k} ]\sum_{m \neq n}\frac{\Omega^2}{(\hbar\Omega -\varepsilon_{mn})^2 +\varsigma^2}  \Delta_{mn}^{\alpha} G_{mn}^{\alpha_1\alpha_2}g_{mn}.
\end{aligned}
\end{equation}
Hence it converges in the dc limit.
The second term in \Eq{eta-ee-d} is
\begin{equation}\label{eta-ee-d-2}
\begin{aligned}
\eta^{\alpha\alpha_1 \alpha_2}_{\text{ee,d,2}}=
&-\frac{\nu_c\hbar}{2 }\int [d \bm{k} ]\sum_{m \neq n}E_{S}^{\alpha_1\alpha_2\beta\gamma}\frac{1}{(\hbar\Omega - \varepsilon_{mn})^2} \Omega^2\Delta_{mn}^{\alpha} \mathcal{A}_{  nm}^{\beta} \mathcal{A}_{  mn}^{\gamma} g_{mn}\\
=&-2\varsigma \nu_c\hbar\int [d \bm{k} ]\sum_{m \neq n}\frac{1}{(\hbar\Omega - \varepsilon_{mn})^2} \Omega^2\Delta_{mn}^{\alpha} G_{mn}^{\alpha_1\alpha_2} g_{mn}\\
=&-2\varsigma \nu_c\hbar\int [d \bm{k} ]\sum_{m \neq n} \partial^{\alpha}\left(\frac{1}{\hbar\Omega - \varepsilon_{mn}}\right)\Omega^2 G_{mn}^{\alpha_1\alpha_2} g_{mn}.
\end{aligned}
\end{equation}
We will show in the following \Eq{eta-ee-d-2} is a part of the LP rectification current.
The CP response of $\chi^{\alpha\alpha_1 \alpha_2}_{\text{ee,d}}$ is
\begin{eqnarray}
\kappa^{\alpha\alpha_1 \alpha_2}_{\text{ee,d}} &=& \lim_{\omega \rightarrow 0} \frac{i\nu_c\hbar}{2 \omega}\int [d \bm{k} ]\sum_{m \neq n}E_{A}^{\alpha_1\alpha_2\beta\gamma}\left(\frac{1}{-\hbar\Omega -\varepsilon_{nm}} - \frac{1}{\hbar\omega + \hbar\Omega - \varepsilon_{nm}}\right) \notag\\
&&\times \Omega(\omega + \Omega)\Delta_{mn}^{\alpha} \mathcal{A}_{  mn}^{\beta} \mathcal{A}_{  nm}^{\gamma}g_{mn} \notag\\
&=& \lim_{\omega \rightarrow 0} \frac{i\nu_c\hbar}{2\omega}\int [d \bm{k} ]\sum_{m \neq n}E_{A}^{\alpha_1\alpha_2\beta\gamma}\left(\frac{1}{-\hbar\Omega -\varepsilon_{nm}} \mathcal{A}_{  mn}^{\beta} \mathcal{A}_{  nm}^{\gamma}- \frac{1}{\hbar\omega + \hbar\Omega + \varepsilon_{nm}} \mathcal{A}_{  nm}^{\beta} \mathcal{A}_{  mn}^{\gamma}\right) \notag\\
&& \times \Omega(\omega + \Omega)\Delta_{mn}^{\alpha} g_{mn}.
\label{kappa-alpha12}
\end{eqnarray}
By use of \Eq{taylor} and \Eq{cauchy}, we have
\begin{equation}\label{kappa-ee-d}
\begin{aligned}
\kappa^{\alpha\alpha_1 \alpha_2}_{\text{ee,d}}=&\lim_{\omega \rightarrow 0} \frac{\pi\nu_c\hbar}{2 \omega}\int [d \bm{k} ]\sum_{m \neq n}E_{A}^{\alpha_1\alpha_2\beta\gamma}\delta({\hbar\Omega -\varepsilon_{mn}}) \Omega(\omega + \Omega)\Delta_{mn}^{\alpha} (\mathcal{A}_{  mn}^{\beta} \mathcal{A}_{  nm}^{\gamma} -\mathcal{A}_{  nm}^{\beta} \mathcal{A}_{  mn}^{\gamma})g_{mn}\\
&-\lim_{\omega \rightarrow 0} \frac{i\nu_c\hbar}{2 \omega}\int [d \bm{k} ]\sum_{m \neq n}E_{A}^{\alpha_1\alpha_2\beta\gamma}\text{P}\frac{1}{\hbar\Omega -\varepsilon_{mn}} \Omega(\omega + \Omega)\Delta_{mn}^{\alpha} (\mathcal{A}_{  mn}^{\beta} \mathcal{A}_{  nm}^{\gamma} +\mathcal{A}_{  nm}^{\beta} \mathcal{A}_{  mn}^{\gamma})g_{mn}\\
&+\lim_{\omega \rightarrow 0} \frac{i\nu_c\hbar}{2 \omega}\int [d \bm{k} ]\sum_{m \neq n}E_{A}^{\alpha_1\alpha_2\beta\gamma}\frac{\omega}{(\hbar\Omega -\varepsilon_{mn})^2} \Omega(\omega + \Omega)\Delta_{mn}^{\alpha} \mathcal{A}_{  mn}^{\beta} \mathcal{A}_{  nm}^{\gamma}g_{mn}.
.\\
\end{aligned}
\end{equation}
The first term in \Eq{kappa-ee-d} is the CP injection current, which is written as
\begin{equation}\label{kappa-shift}
\begin{aligned}
\kappa^{\alpha\alpha_1 \alpha_2}_{\text{Inj}}=&\lim_{\omega \rightarrow 0} \frac{\pi\nu_c\hbar}{2 \omega}\int [d \bm{k} ]\sum_{m \neq n}E_{A}^{\alpha_1\alpha_2\beta\gamma}\delta({\hbar\Omega -\varepsilon_{mn}}) \Omega(\omega + \Omega)\Delta_{mn}^{\alpha} (\mathcal{A}_{  mn}^{\beta} \mathcal{A}_{  nm}^{\gamma} -\mathcal{A}_{  nm}^{\beta} \mathcal{A}_{  mn}^{\gamma})g_{mn}\\
=&-\lim_{\omega \rightarrow 0} \frac{i\pi\nu_c\hbar}{2 \omega}\int [d \bm{k} ]\sum_{m \neq n}E_{A}^{\alpha_1\alpha_2\beta\gamma}\delta({\hbar\Omega -\varepsilon_{mn}}) \Omega(\omega + \Omega)\Delta_{mn}^{\alpha} \Omega_{mn}^{\beta\gamma} g_{mn}\\
=&- \frac{\pi\nu_c\hbar}{2 }\int [d \bm{k} ]\sum_{m \neq n}E_{A}^{\alpha_1\alpha_2\beta\gamma}\frac{\Omega^2}{(\hbar\Omega -\varepsilon_{mn})^2 +\varsigma^2} \Delta_{mn}^{\alpha} \Omega_{mn}^{\beta\gamma} g_{mn}\\
=&- {\pi\nu_c\hbar}\int [d \bm{k} ]\sum_{m \neq n}\frac{\Omega^2}{(\hbar\Omega -\varepsilon_{mn})^2 +\varsigma^2} \Delta_{mn}^{\alpha} \Omega_{mn}^{\alpha_1\alpha_2} g_{mn}.
\end{aligned}
\end{equation}
The second term in \Eq{kappa-ee-d} is
\begin{equation}
\begin{aligned}
\kappa^{\alpha\alpha_1 \alpha_2}_{\text{ee,d,2}}=&-\lim_{\omega \rightarrow 0} \frac{i\nu_c\hbar}{\omega}\int [d \bm{k} ]\sum_{m \neq n}E_{A}^{\alpha_1\alpha_2\beta\gamma}\text{P}\frac{1}{\hbar\Omega -\varepsilon_{mn}} \Omega^2 \Delta_{mp}^{\alpha} G_{mn}^{\beta\gamma}g_{mn}.
\end{aligned}
\end{equation}
Noting that in the CP responses $\beta \neq \gamma$, $E_{A}^{\alpha_1\alpha_2\beta\gamma} = - E_{A}^{\alpha_1\alpha_2\gamma\beta}$, and making use of the fact $G_{mn}^{\beta\gamma} = G_{mn}^{\gamma\beta}$, we have $\kappa^{\alpha\alpha_1 \alpha_2}_{\text{ee,d,2}} = 0$.
 The third term in \Eq{kappa-ee-d} is written as
\begin{equation}\label{kappa-ee-d-2}
\begin{aligned}
\kappa^{\alpha\alpha_1 \alpha_2}_{\text{Rec,1}}=&\frac{i\nu_c\hbar}{2 \omega}\int [d \bm{k} ]\sum_{m \neq n}\frac{\omega}{(\hbar\Omega -\varepsilon_{mn})^2} \Omega^2\Delta_{mp}^{\alpha} (\mathcal{A}_{  mn}^{\alpha_1} \mathcal{A}_{  nm}^{\alpha_2} - \mathcal{A}_{  mn}^{\alpha_2} \mathcal{A}_{  nm}^{\alpha_1})g_{mn}\\
=&\frac{\nu_c\hbar}{2 }\int [d \bm{k} ]\sum_{m \neq n}\frac{\Omega^2}{(\hbar\Omega -\varepsilon_{mn})^2} \Delta_{mp}^{\alpha} \Omega_{mn}^{\alpha_1\alpha_2} g_{mn}\\
=&-\frac{\nu_c\hbar}{2 }\int [d \bm{k} ]\sum_{m \neq n}\partial^{\alpha}\left(\frac{1}{\hbar\Omega -\varepsilon_{mn}}\right) \Omega^2 \Omega_{mn}^{\alpha_1\alpha_2} g_{mn}.
\end{aligned}
\end{equation}
We will show in the following $\kappa^{\alpha\alpha_1 \alpha_2}_{\text{Rec,1}}$ is a part of the CP rectification current.

\subsection{Shift spin current }
Now we consider the remaining terms $\chi_{\text{ie}}$ and the nondiagonal component of $\chi_{\text{ee}}$ which is noted as $\chi_{\text{ee,od}}$. We show that the summation of these two terms comprises the shift current.
By use of used partial integration, $\chi_{\text{ie}}$ can be rewritten as
\begin{eqnarray}
\chi^{\alpha\alpha_1 \alpha_2}_{\text{ie}} &=& \lim_{\omega \rightarrow 0}\frac{\nu_c}{2}\int [d \bm{k} ]\sum_{m\neq n} \epsilon^{\alpha_1 z \beta}\epsilon^{\alpha_2 z \gamma}{\omega_1\omega_2}\left[-\partial^{\beta}\mathcal{A}_{mn}^{\alpha} + i(\mathcal{A}^{\beta}_{nn}-\mathcal{A}^{\beta}_{mm})\mathcal{A}_{mn}^{\alpha}\right] \notag\\
&& \times d_{nm}(\omega_2)  \mathcal{A}^{\gamma}_{nm}g_{nm}+ [(\alpha_1 , \omega_1)\leftrightarrow (\alpha_2, \omega_2)] \notag\\
&=& \lim_{\omega \rightarrow 0}\frac{\nu_c}{2}\int [d \bm{k} ]\sum_{m\neq n} \left(\epsilon^{\alpha_1 z \beta}\epsilon^{\alpha_2 z \gamma}\frac{1}{-\hbar\Omega -\varepsilon_{nm}} + \epsilon^{\alpha_2 z \beta}\epsilon^{\alpha_1 z \gamma}\frac{1}{\hbar\omega + \hbar\Omega - \varepsilon_{nm}}\right){\omega_1\omega_2} \notag\\
&&\times\left[-\partial^{\beta}\mathcal{A}_{mn}^{\alpha} + i(\mathcal{A}^{\beta}_{nn}-\mathcal{A}^{\beta}_{mm})\mathcal{A}_{mn}^{\alpha}\right]  \mathcal{A}^{\gamma}_{nm}g_{nm}.
\end{eqnarray}
Noting that
\begin{equation}
\partial^{\beta}\mathcal{A}_{mn}^{\alpha} - i(\mathcal{A}^{\beta}_{nn}-\mathcal{A}^{\beta}_{mm})\mathcal{A}_{mn}^{\alpha} = [D^{\beta}\mathcal{A}^{\alpha}]_{mn},
\end{equation}
Making use of \Eq{taylor} and \Eq{cauchy}, the LP-photocurrent tensor is obtained as
\begin{equation}\label{eta-ie}
\begin{aligned}
\eta^{\alpha\alpha_1 \alpha_2}_{\text{ie}}=&-\lim_{\omega \rightarrow 0}\frac{\nu_c}{2}\int [d \bm{k} ]\sum_{m\neq n} E_{S}^{\alpha_1\alpha_2\beta\gamma}\left(\frac{1}{-\hbar\Omega -\varepsilon_{nm}}[D^{\beta}\mathcal{A}^{\alpha}]_{mn}  \mathcal{A}^{\gamma}_{nm} \right. \\
& - \left.\frac{1}{\hbar\omega + \hbar\Omega + \varepsilon_{nm}}[D^{\beta}\mathcal{A}^{\alpha}]_{nm}  \mathcal{A}^{\gamma}_{mn}\right)\Omega(\omega + \Omega)g_{nm}\\
=&\frac{\nu_c}{2}\int [d \bm{k} ]\sum_{m\neq n} E_{S}^{\alpha_1\alpha_2\beta\gamma}\text{P}\frac{1}{\hbar\Omega -\varepsilon_{mn}}\Omega^2 \left([D^{\beta}\mathcal{A}^{\alpha}]_{mn}  \mathcal{A}^{\gamma}_{nm} + [D^{\beta}\mathcal{A}^{\alpha}]_{nm}  \mathcal{A}^{\gamma}_{mn}\right)g_{nm}\\
&+i\frac{\pi\nu_c}{2}\int [d \bm{k} ]\sum_{m\neq n} E_{S}^{\alpha_1\alpha_2\beta\gamma}\delta(\hbar\Omega -\varepsilon_{mn})\Omega^2 \left([D^{\beta}\mathcal{A}^{\alpha}]_{mn}  \mathcal{A}^{\gamma}_{nm} - [D^{\beta}\mathcal{A}^{\alpha}]_{nm}  \mathcal{A}^{\gamma}_{mn}\right)g_{nm}.
\end{aligned}
\end{equation}

The CP-photocurrent tensor is
\begin{equation}\label{kappa-ie}
\begin{aligned}
\kappa^{\alpha\alpha_1 \alpha_2}_{\text{ie}}=&\lim_{\omega \rightarrow 0}\frac{i\nu_c}{2}\int [d \bm{k} ]\sum_{m\neq n} E_{A}^{\alpha_1\alpha_2\beta\gamma}\left(\frac{1}{-\hbar\Omega -\varepsilon_{nm}}[D^{\beta}\mathcal{A}^{\alpha}]_{mn}  \mathcal{A}^{\gamma}_{nm}\right.  \\
& + \left.\frac{1}{\hbar\omega + \hbar\Omega + \varepsilon_{nm}}[D^{\beta}\mathcal{A}^{\alpha}]_{nm}  \mathcal{A}^{\gamma}_{mn}\right)\Omega(\omega + \Omega)g_{nm}\\
=&\frac{\pi\nu_c}{2}\int [d \bm{k} ]\sum_{m\neq n} E_{A}^{\alpha_1\alpha_2\beta\gamma}\delta(\hbar\Omega -\varepsilon_{mn})\Omega^2 \left([D^{\beta}\mathcal{A}^{\alpha}]_{mn}  \mathcal{A}^{\gamma}_{nm} + [D^{\beta}\mathcal{A}^{\alpha}]_{nm}  \mathcal{A}^{\gamma}_{mn}\right)g_{nm}\\
&-\frac{i\nu_c}{2}\int [d \bm{k} ]\sum_{m\neq n} E_{A}^{\alpha_1\alpha_2\beta\gamma}\text{P}\frac{1}{\hbar\Omega -\varepsilon_{mn}}\Omega^2 \left([D^{\beta}\mathcal{A}^{\alpha}]_{mn}  \mathcal{A}^{\gamma}_{nm} - [D^{\beta}\mathcal{A}^{\alpha}]_{nm}  \mathcal{A}^{\gamma}_{mn}\right)g_{nm}.
\end{aligned}
\end{equation}

Now we consider the remaining component of $\chi_{\text{ee}}$, which is denoted as $\chi_{\text{ee,od}}$.
\begin{equation}
\begin{aligned}
\chi^{\alpha\alpha_1 \alpha_2}_{\text{ee,od}}= & \lim_{\omega\rightarrow 0}i\frac{\nu_c}{2}\int [d \bm{k} ]\sum_{m\neq n\neq p}\epsilon^{\alpha_1 z \beta}\epsilon^{\alpha_2 z \gamma}{\omega_1\omega_2}\mathcal{A}_{mn}^{\alpha} \left[d_{pm}(\omega_2)\mathcal{A}_{  np}^{\beta}\mathcal{A}_{  pm}^{\gamma}g_{mp}  -d_{np}(\omega_2)\mathcal{A}_{  pm}^{\beta}\mathcal{A}_{  np}^{\gamma}g_{pn}\right]\\
&+ [(\alpha_1 , \omega_1)\leftrightarrow (\alpha_2, \omega_2)]\\
= & \lim_{\omega\rightarrow 0}i\frac{\nu_c}{2}\int [d \bm{k} ]\sum_{m\neq n\neq p}\epsilon^{\alpha_1 z \beta}\epsilon^{\alpha_2 z \gamma}{\omega_1\omega_2}(\mathcal{A}_{mn}^{\alpha}\mathcal{A}_{  np}^{\beta}- \mathcal{A}_{mn}^{\beta}\mathcal{A}_{  np}^{\alpha})d_{pm}(\omega_2)\mathcal{A}_{  pm}^{\gamma}g_{mp}  \\
& + [(\alpha_1 , \omega_1)\leftrightarrow (\alpha_2, \omega_2)]\\
= & \lim_{\omega\rightarrow 0}i\frac{\nu_c}{2}\int [d \bm{k} ]\sum_{m\neq n\neq p}\left(\epsilon^{\alpha_1 z \beta}\epsilon^{\alpha_2 z \gamma}\frac{1}{-\hbar\Omega -\varepsilon_{pm}} + \epsilon^{\alpha_2 z \beta}\epsilon^{\alpha_1 z \gamma}\frac{1}{\hbar\omega + \hbar\Omega - \varepsilon_{pm}}\right) \\
&\times \Omega(\omega + \Omega)(\mathcal{A}_{mn}^{\alpha}\mathcal{A}_{  np}^{\beta}- \mathcal{A}_{mn}^{\beta}\mathcal{A}_{  np}^{\alpha})\mathcal{A}_{  pm}^{\gamma}g_{mp}\\
= & \lim_{\omega\rightarrow 0}\frac{\nu_c}{2}\int [d \bm{k} ]\sum_{m\neq p}\left(\epsilon^{\alpha_1 z \beta}\epsilon^{\alpha_2 z \gamma}\frac{1}{-\hbar\Omega -\varepsilon_{pm}} + \epsilon^{\alpha_2 z \beta}\epsilon^{\alpha_1 z \gamma}\frac{1}{\hbar\omega + \hbar\Omega - \varepsilon_{pm}}\right)\Omega(\omega + \Omega)\\
&\times\left([D^{\alpha}\mathcal{A}^{\beta}]_{mp} - [D^{\beta}\mathcal{A}^{\alpha}]_{mp}\right)\mathcal{A}_{  pm}^{\gamma}g_{mp}.
\end{aligned}
\end{equation}
Where in the last equality the relation $\sum_{n\neq m,p}(\mathcal{A}_{mn}^{\alpha}\mathcal{A}_{  np}^{\beta}- \mathcal{A}_{mn}^{\beta}\mathcal{A}_{  np}^{\alpha})=-i([D^{\alpha}\mathcal{A}^{\beta}]_{mp} - [D^{\beta}\mathcal{A}^{\alpha}]_{mp})$ is used. For notation simplicity, we note $K_{mp}^{\alpha\beta\gamma}=\left([D^{\alpha}\mathcal{A}^{\beta}]_{mp} - [D^{\beta}\mathcal{A}^{\alpha}]_{mp}\right)\mathcal{A}_{  pm}^{\gamma}$,
the LP response of $\chi^{\alpha\alpha_1 \alpha_2}_{\text{ee,od}}$ is written as
\begin{equation}\label{eta-ee-od}
\begin{aligned}
\eta^{\alpha\alpha_1 \alpha_2}_{\text{ee,od}}
= & \lim_{\omega\rightarrow 0}\frac{\nu_c}{2}\int [d \bm{k} ]\sum_{m\neq p} E_{S}^{\alpha_1\alpha_2\beta\gamma}\left(\frac{1}{-\hbar\Omega -\varepsilon_{pm}} + \frac{1}{\hbar\omega + \hbar\Omega - \varepsilon_{pm}}\right)\Omega(\omega + \Omega)K_{mp}^{\alpha\beta\gamma}g_{mp}\\
=&\lim_{\omega\rightarrow 0}\frac{\nu_c}{2}\int [d \bm{k} ]\sum_{m\neq p} E_{S}^{\alpha_1\alpha_2\beta\gamma}\left(\frac{1}{-\hbar\Omega -\varepsilon_{pm}}K_{mp}^{\alpha\beta\gamma} - \frac{1}{\hbar\omega + \hbar\Omega + \varepsilon_{pm}}K_{pm}^{\alpha\beta\gamma}\right)\Omega^2g_{mp}\\
=&-\frac{\nu_c}{2}\int [d \bm{k} ]\sum_{m\neq p} E_{S}^{\alpha_1\alpha_2\beta\gamma}\text{P}\frac{1}{\hbar\Omega -\varepsilon_{mp}}\left(K_{mp}^{\alpha\beta\gamma} +K_{pm}^{\alpha\beta\gamma}\right)\Omega^2g_{mp}\\
&-i\frac{\pi\nu_c}{2}\int [d \bm{k} ]\sum_{m\neq p} E_{S}^{\alpha_1\alpha_2\beta\gamma}\delta(\hbar\Omega -\varepsilon_{mp})\left(K_{mp}^{\alpha\beta\gamma} -K_{pm}^{\alpha\beta\gamma}\right)\Omega^2g_{mp}.
\end{aligned}
\end{equation}
Adding \Eq{eta-ie} and \Eq{eta-ee-od} results to:
\begin{equation}
\begin{aligned}
&\eta^{\alpha\alpha_1 \alpha_2}_{\text{ie}} + \eta^{\alpha\alpha_1 \alpha_2}_{\text{ee,od}}\\
=&-i \frac{\pi \nu_c}{2}\int [d \bm{k} ]\sum_{m\neq n}E_{S}^{\alpha_1\alpha_2\beta\gamma}\Omega^2 \delta(\hbar\Omega -\varepsilon_{mn})\left([D^{\alpha}\mathcal{A}^{\beta}]_{mn} \mathcal{A}_{  nm}^{\gamma} - [D^{\alpha}\mathcal{A}^{\beta}]_{nm}\mathcal{A}_{mn}^{\gamma}\right)g_{mn}\\
&- \frac{ \nu_c}{2}\int [d \bm{k} ]\sum_{m\neq n}E_{S}^{\alpha_1\alpha_2\beta\gamma}\Omega^2 \text{P}\frac{1}{\hbar\Omega - \varepsilon_{mn}}\left([D^{\alpha}\mathcal{A}^{\beta}]_{mn} \mathcal{A}_{  nm}^{\gamma} + [D^{\alpha}\mathcal{A}^{\beta}]_{nm}\mathcal{A}_{mn}^{\gamma}\right)g_{mn}\\
=&-i\varsigma \frac{\pi \nu_c}{2}\int [d \bm{k} ]\sum_{m\neq n}\Omega^2 \delta(\hbar\Omega -\varepsilon_{mn})\left([D^{\alpha}\mathcal{A}^{\alpha_1}]_{mn} \mathcal{A}_{  nm}^{\alpha_2} - [D^{\alpha}\mathcal{A}^{\alpha_1}]_{nm}\mathcal{A}_{mn}^{\alpha_2} \right. \\
& + \left.[D^{\alpha}\mathcal{A}^{\alpha_2}]_{mn} \mathcal{A}_{  nm}^{\alpha_1} - [D^{\alpha}\mathcal{A}^{\alpha_2}]_{nm}\mathcal{A}_{mn}^{\alpha_1}\right)g_{mn}\\
& -\varsigma\frac{ \nu_c}{2}\int [d \bm{k} ]\sum_{m\neq n}\Omega^2 \text{P}\frac{1}{\hbar\Omega - \varepsilon_{mn}}\left([D^{\alpha}\mathcal{A}^{\alpha_1}]_{mn} \mathcal{A}_{  nm}^{\alpha_2} + [D^{\alpha}\mathcal{A}^{\alpha_1}]_{nm}\mathcal{A}_{mn}^{\alpha_2} \right. \\
& + \left. [D^{\alpha}\mathcal{A}^{\alpha_2}]_{mn} \mathcal{A}_{  nm}^{\alpha_1} + [D^{\alpha}\mathcal{A}^{\alpha_2}]_{nm}\mathcal{A}_{mn}^{\alpha_1}\right)g_{mn}.
\end{aligned}
\end{equation}
Making use of the relation
\begin{equation}\label{relation1}
[D^{\alpha}\mathcal{A}^{\alpha_1}]_{mn} \mathcal{A}_{  nm}^{\alpha_2} = ([D^{\alpha}\mathcal{A}^{\alpha_1}]_{nm} \mathcal{A}_{  mn}^{\alpha_2})^*,
\end{equation}
 we have
\begin{equation}\label{eta-ie-eeod}
\begin{aligned}
\eta^{\alpha\alpha_1 \alpha_2}_{\text{ie}} & + \eta^{\alpha\alpha_1 \alpha_2}_{\text{ee,od}} \\
&= \varsigma\pi \nu_c\int [d \bm{k} ]\sum_{m\neq n}\Omega^2 \delta(\hbar\Omega -\varepsilon_{mn})\text{Im}\left([D^{\alpha}\mathcal{A}^{\alpha_1}]_{mn} \mathcal{A}_{  nm}^{\alpha_2}  + [D^{\alpha}\mathcal{A}^{\alpha_2}]_{mn} \mathcal{A}_{  nm}^{\alpha_1} \right)g_{mn}\\
& -\varsigma\nu_c\int [d \bm{k} ]\sum_{m\neq n}\Omega^2 \text{P}\frac{1}{\hbar\Omega - \varepsilon_{mn}}\text{Re}\left([D^{\alpha}\mathcal{A}^{\alpha_1}]_{mn} \mathcal{A}_{  nm}^{\alpha_2}+ [D^{\alpha}\mathcal{A}^{\alpha_2}]_{mn} \mathcal{A}_{  nm}^{\alpha_1}\right)g_{mn}.
\end{aligned}
\end{equation}
The first term in \Eq{eta-ie-eeod} is the shift current.   Following Ref. \cite{PhysRevB.61.5337, PhysRevX.10.041041}, we define the shift vector:
\begin{equation}\label{shift-vec}
R_{mn}^{\alpha_1\alpha_2} = \mathcal{A}_{mm}^{\alpha_1} - \mathcal{A}_{nn}^{\alpha_1} - \partial^{\alpha_1} \arg \mathcal{A}_{mn}^{\alpha_2}.
\end{equation}
We can write the LP-shift current in terms of the magnon shift vector as
\begin{equation}\label{shift-lp}
\begin{aligned}
\eta^{\alpha\alpha_1 \alpha_2}_{\text{Sh}}=& \frac{\pi \nu_c}{2}\int [d \bm{k} ]\sum_{m\neq n}\Omega^2 \delta(\hbar\Omega -\varepsilon_{mn})(R_{mn}^{\alpha\alpha_1}+R_{mn}^{\alpha\alpha_2}) G_{mn}^{\alpha_1\alpha_2} g_{mn}\\
&-\frac{\pi \nu_c}{2}\int [d \bm{k} ]\sum_{m\neq n}\Omega^2 \delta(\hbar\Omega -\varepsilon_{mn}) (\lvert \mathcal{A}_{nm}^{\alpha_2}\rvert \partial^{\alpha}\lvert \mathcal{A}_{mn}^{\alpha_1}\rvert - \lvert \mathcal{A}_{mn}^{\alpha_1}\rvert \partial^{\alpha}\lvert \mathcal{A}_{nm}^{\alpha_2} \rvert) \\
&\times \sin{(\phi_{mn}^{\alpha_1}+ \phi_{nm}^{\alpha_2})} g_{mn}.
\end{aligned}
\end{equation}
It describes the current generate by the shift of the magnon position in the inter-band transition from band $m$ to $n$. One can verify that the shift current is invariant to the gauge transformation. By use of the relation
\begin{equation}\label{relation2}
\text{Re}\left([D^{\alpha}\mathcal{A}^{\alpha_1}]_{mn} \mathcal{A}_{  nm}^{\alpha_2}+ [D^{\alpha}\mathcal{A}^{\alpha_2}]_{mn} \mathcal{A}_{  nm}^{\alpha_1}\right) = \partial^{\alpha}G_{mn}^{\alpha_1\alpha_2}
\end{equation}
The second term in \Eq{eta-ie-eeod} is a part of the rectification current, which is written as
\begin{equation}\label{eta-ie-eeod-2}
\begin{aligned}
 \eta^{\alpha\alpha_1 \alpha_2}_{\text{Rec,2}}=&-\varsigma 2 \nu_c\int [d \bm{k} ]\sum_{m\neq n}\Omega^2 \text{P}\frac{1}{\hbar\Omega - \varepsilon_{mn}}\partial^{\alpha}G_{mn}^{\alpha_1\alpha_2}g_{mn}.
\end{aligned}
\end{equation}
Combining \Eq{eta-ee-d-2} and \Eq{eta-ie-eeod-2}, we have
\begin{equation}
\begin{aligned}
\eta^{\alpha\alpha_1 \alpha_2}_{\text{Rec}}= \eta^{\alpha\alpha_1 \alpha_2}_{\text{Rec,1}} + \eta^{\alpha\alpha_1 \alpha_2}_{\text{Rec,2}} =& 2 \nu_c\int [d \bm{k} ]\sum_{m\neq n}\Omega^2 \frac{1}{\hbar\Omega - \varepsilon_{mn}}G_{mn}^{\alpha_1\alpha_2}\partial^{\alpha}g_{mn}.
\end{aligned}
\end{equation}

The CP-photocurrent tensor is
\begin{equation}\label{kappa-ee-od}
\begin{aligned}
\kappa^{\alpha\alpha_1 \alpha_2}_{\text{ee,od}}
= & \lim_{\omega\rightarrow 0}i\frac{\nu_c}{2}\int [d \bm{k} ]\sum_{m\neq p} E_{A}^{\alpha_1\alpha_2\beta\gamma}\left(\frac{1}{-\hbar\Omega -\varepsilon_{pm}} - \frac{1}{\hbar\omega + \hbar\Omega - \varepsilon_{pm}}\right) \\
&\times \Omega(\omega + \Omega)\left([D^{\alpha}\mathcal{A}^{\beta}]_{mp} - [D^{\beta}\mathcal{A}^{\alpha}]_{mp}\right)\mathcal{A}_{pm}^{\gamma}g_{mp}\\
=&\lim_{\omega\rightarrow 0}i\frac{\nu_c}{2}\int [d \bm{k} ]\sum_{m\neq p} E_{S}^{\alpha_1\alpha_2\beta\gamma}\left(\frac{1}{-\hbar\Omega -\varepsilon_{pm}}K_{mp}^{\alpha\beta\gamma} + \frac{1}{\hbar\omega + \hbar\Omega + \varepsilon_{pm}}K_{pm}^{\alpha\beta\gamma}\right)\Omega^2g_{mp}\\
=&-i\frac{\nu_c}{2}\int [d \bm{k} ]\sum_{m\neq p} E_{S}^{\alpha_1\alpha_2\beta\gamma}\text{P}\frac{1}{\hbar\Omega -\varepsilon_{mn}}\left(K_{mp}^{\alpha\beta\gamma} -K_{pm}^{\alpha\beta\gamma}\right)\Omega^2g_{mp}\\
&+\frac{\pi\nu_c}{2}\int [d \bm{k} ]\sum_{m\neq p} E_{A}^{\alpha_1\alpha_2\beta\gamma}\delta(\hbar\Omega -\varepsilon_{mn})\left[K_{mp}^{\alpha\beta\gamma} +K_{pm}^{\alpha\beta\gamma}\right]\Omega^2g_{mp}.
\end{aligned}
\end{equation}
Adding \Eq{kappa-ie} and \Eq{kappa-ee-od}, we have
\begin{equation}
\begin{aligned}
&\kappa^{\alpha\alpha_1 \alpha_2}_{\text{ee,od}}+\kappa^{\alpha\alpha_1 \alpha_2}_{\text{ie}}\\
=&-i\frac{\nu_c}{2}\int [d \bm{k} ]\sum_{m\neq p} E_{S}^{\alpha_1\alpha_2\beta\gamma}\text{P}\frac{1}{\hbar\Omega -\varepsilon_{mn}}\left([D^{\alpha}\mathcal{A}^{\beta}]_{mp} \mathcal{A}_{pm}^{\gamma} -[D^{\alpha}\mathcal{A}^{\beta}]_{pm} \mathcal{A}_{mp}^{\gamma}\right)\Omega^2g_{mp}\\
&+\frac{\pi\nu_c}{2}\int [d \bm{k} ]\sum_{m\neq p} E_{A}^{\alpha_1\alpha_2\beta\gamma}\delta(\hbar\Omega -\varepsilon_{mn})\left([D^{\alpha}\mathcal{A}^{\beta}]_{mp} \mathcal{A}_{pm}^{\gamma} +[D^{\alpha}\mathcal{A}^{\beta}]_{pm} \mathcal{A}_{mp}^{\gamma}\right)\Omega^2g_{mp}\\
=&-\frac{\pi\nu_c}{2}\int [d \bm{k} ]\sum_{m\neq p} \delta(\hbar\Omega -\varepsilon_{mn})\left([D^{\alpha}\mathcal{A}^{\alpha_1}]_{mp} \mathcal{A}_{pm}^{\alpha_2} +[D^{\alpha}\mathcal{A}^{\alpha_1}]_{pm} \mathcal{A}_{mp}^{\alpha_2} \right. \\
&-\left.[D^{\alpha}\mathcal{A}^{\alpha_2}]_{mp} \mathcal{A}_{pm}^{\alpha_1} -[D^{\alpha}\mathcal{A}^{\alpha_2}]_{pm} \mathcal{A}_{mp}^{\alpha_1}\right)\Omega^2g_{mp}\\
&+i\frac{\nu_c}{2}\int [d \bm{k} ]\sum_{m\neq p} \text{P}\frac{1}{\hbar\Omega -\varepsilon_{mn}}\left([D^{\alpha}\mathcal{A}^{\alpha_1}]_{mp} \mathcal{A}_{pm}^{\alpha_2} \right. \\ 
&-\left.[D^{\alpha}\mathcal{A}^{\alpha_1}]_{pm} \mathcal{A}_{mp}^{\alpha_2} -[D^{\alpha}\mathcal{A}^{\alpha_2}]_{mp} \mathcal{A}_{pm}^{\alpha_1} +[D^{\alpha}\mathcal{A}^{\alpha_2}]_{pm} \mathcal{A}_{mp}^{\alpha_1}\right)\Omega^2g_{mp}.
\end{aligned}
\end{equation}
Using \Eq{relation1} and \Eq{relation2}, we have
\begin{equation}\label{kappa-ie-eeod}
\begin{aligned}
\kappa^{\alpha\alpha_1 \alpha_2}_{\text{ee,od}} & +\kappa^{\alpha\alpha_1 \alpha_2}_{\text{ie}} \\
&=-\frac{\pi\nu_c}{2}\int [d \bm{k} ]\sum_{m\neq p} \delta(\hbar\Omega -\varepsilon_{mn})\text{Re}\left([D^{\alpha}\mathcal{A}^{\alpha_1}]_{mp} \mathcal{A}_{pm}^{\alpha_2} -[D^{\alpha}\mathcal{A}^{\alpha_2}]_{mp} \mathcal{A}_{pm}^{\alpha_1} \right)\Omega^2g_{mp}\\
&-\frac{\nu_c}{2}\int [d \bm{k} ]\sum_{m\neq p} \text{P}\frac{1}{\hbar\Omega -\varepsilon_{mn}}\text{Im}\left([D^{\alpha}\mathcal{A}^{\alpha_1}]_{mp} \mathcal{A}_{pm}^{\alpha_2} -[D^{\alpha}\mathcal{A}^{\alpha_2}]_{mp} \mathcal{A}_{pm}^{\alpha_1} \right)\Omega^2g_{mp}.
\end{aligned}
\end{equation}
Inheriting the concept from the electron photocurrent responses, the first term in \Eq{kappa-ie-eeod} is the CP shift current:
\begin{equation}\label{kappa-shift-1}
\begin{aligned}
\kappa^{\alpha\alpha_1 \alpha_2}_{\text{Sh}}=-\frac{\pi\nu_c}{2}\int [d \bm{k} ]\sum_{m\neq p} \delta(\hbar\Omega -\varepsilon_{mn})\text{Re}\left([D^{\alpha}\mathcal{A}^{\alpha_1}]_{mp} \mathcal{A}_{pm}^{\alpha_2} -[D^{\alpha}\mathcal{A}^{\alpha_2}]_{mp} \mathcal{A}_{pm}^{\alpha_1} \right)\Omega^2g_{mp}.
\end{aligned}
\end{equation}
Introducing the chiral shift vector:
\begin{equation}
R_{mn}^{\alpha, \pm} = \mathcal{A}_{mm}^{\alpha} - \mathcal{A}_{nn}^{\alpha} - \partial^{\alpha} \arg \mathcal{A}_{mn}^{\pm},
\end{equation}
by use of the circular representation of the Berry connection $\mathcal{A}_{mn}^{\pm} = \frac{1}{\sqrt{2}}(\mathcal{A}_{mn}^{x}\pm i\mathcal{A}_{mn}^{y})$, \Eq{kappa-shift-1} can be rewritten as
\begin{equation}
\kappa^{\alpha\alpha_1 \alpha_2}_{\text{Sh}}=-\frac{\pi\nu_c}{2}\int [d \bm{k} ]\sum_{m\neq n} \Omega^2\delta(\hbar\Omega -\varepsilon_{mn})(R_{mn}^{\alpha, +}\lvert\mathcal{A}_{nm}^{+}\rvert^2 -R_{mn}^{\alpha, -}\lvert\mathcal{A}_{nm}^{-}\rvert^2) g_{mn}.
\end{equation}
By use of the relation
\begin{equation}
\text{Im}\left([D^{\alpha}\mathcal{A}^{\alpha_1}]_{mp} \mathcal{A}_{pm}^{\alpha_2} -[D^{\alpha}\mathcal{A}^{\alpha_2}]_{mp} \mathcal{A}_{pm}^{\alpha_1} \right)=-\frac{1}{2}\partial^{\alpha}\Omega_{mn}^{\alpha_1\alpha_2}
\end{equation}
The second term in \Eq{kappa-ie-eeod} is a part of the CP rectification current, which is written as
\begin{equation}\label{kappa-ie-eeod-2}
\begin{aligned}
\kappa^{\alpha\alpha_1 \alpha_2}_{\text{Rec,2}}=\frac{\nu_c}{2}\int [d \bm{k} ]\sum_{m\neq n} \text{P}\frac{1}{\hbar\Omega -\varepsilon_{mn}}\Omega^2\partial^{\alpha}\Omega_{mn}^{\alpha_1\alpha_2}g_{mn}.
\end{aligned}
\end{equation}
Combining \Eq{kappa-ee-d-2} and \Eq{kappa-ie-eeod-2}, we have
\begin{equation}
\begin{aligned}
\kappa^{\alpha\alpha_1 \alpha_2}_{\text{Rec}} = \kappa^{\alpha\alpha_1 \alpha_2}_{\text{Rec,1}}+ \kappa^{\alpha\alpha_1 \alpha_2}_{\text{Rec,2}}=\frac{\nu_c}{2}\int [d \bm{k} ]\sum_{m\neq n} \frac{\Omega^2}{\hbar\Omega -\varepsilon_{mn}}\Omega_{mn}^{\alpha_1\alpha_2}\partial^{\alpha}g_{mn}.
\end{aligned}
\end{equation}

In conclusion, the magnon spin photocurrent is expressed in terms of the following gauge invariant quantities:
\begin{align}
&v_{m}^{\alpha}=\frac{1}{\hbar}\partial^{\alpha}\varepsilon_{m}; \quad \Delta_{mn}^{\alpha}=v_{m}^{\alpha} - v_{n}^{\alpha},  \label{VmDelta} \\
&\Omega_{m}^{\tau} = \frac{i}{2}\epsilon^{\alpha\gamma\tau}\sum_{n\neq m}(\mathcal{A}_{mn}^{\alpha} \mathcal{A}_{  nm}^{\gamma}-\mathcal{A}_{mn}^{\gamma} \mathcal{A}_{  nm}^{\alpha}), \label{Omegamtau}\\
&\Omega_{mn}^{\beta\gamma} =i( \mathcal{A}_{  mn}^{\beta} \mathcal{A}_{  nm}^{\gamma } - \mathcal{A}_{  mn}^{\gamma} \mathcal{A}_{  nm}^{\beta}), \label{Omegamnbetagamma}\\
&G_{mn}^{\beta\gamma} =\frac{1}{2}( \mathcal{A}_{  mn}^{\beta} \mathcal{A}_{  nm}^{\gamma } + \mathcal{A}_{  mn}^{\gamma} \mathcal{A}_{  nm}^{\beta}),\\
&R_{mn}^{\alpha \alpha_1} = \mathcal{A}_{mm}^{\alpha} - \mathcal{A}_{nn}^{\alpha} - \partial^{\alpha} \arg \mathcal{A}_{mn}^{\alpha_1},\\
&R_{mn}^{\alpha, \pm} = \mathcal{A}_{mm}^{\alpha} - \mathcal{A}_{nn}^{\alpha} - \partial^{\alpha} \arg \mathcal{A}_{mn}^{\pm}.
\end{align}
In which $v_{m}^{\alpha}$ is the group velocity, $\Omega_{m}^{\tau}$ is the Berry curvature for the $m$th band,  $\Omega_{mn}^{\beta\gamma}$ is the band-resolved Berry curvature, $G_{mn}^{\beta\gamma}$ is the band-resolved quantum metric, and $R_{mn}^{\alpha_1\alpha_2}$ is the shift vector, $R_{mn}^{\alpha,\pm}$ is the chiral shift vector with the Berry connection in circular representation $\mathcal{A}_{mn}^{\pm} = \frac{1}{\sqrt{2}}(\mathcal{A}_{mn}^{x}\pm i\mathcal{A}_{mn}^{y})$. the magnon spin photocurrent is composed of five distinct parts: Drude current, Berry curvature dipole (BCD) current, injection current, shift current, and the rectification current:
\begin{align}
\eta^{\alpha\alpha_1 \alpha_2}_{\text{D}} =& \varsigma\frac{\nu_c}{\hbar } \int [d \bm{k} ]\sum_m   v_{m}^{\alpha}\partial^{\alpha_1}\partial^{\alpha_2}g_{  m}, \label{eta-drude-o}\\
\kappa^{\alpha\alpha_1 \alpha_2}_{\text{BCD}}=&\frac{\nu_c \Omega}{2\hbar} \int  [d \bm{k} ]\sum_{m} (\epsilon^{\alpha\alpha_2\tau}\partial^{\alpha_1} - \epsilon^{\alpha\alpha_1\tau}\partial^{\alpha_2} )\Omega_{m}^{\tau}g_{m},\label{kappa-bcd-o}\\
 \eta^{\alpha\alpha_1 \alpha_2}_{\text{Inj}}=&-\varsigma {2\hbar\nu_c}\int [d \bm{k} ]\sum_{m \neq n}\frac{\Omega^2}{(\hbar\Omega -\varepsilon_{mn})^2 +\varsigma^2}  \Delta_{mn}^{\alpha} G_{mn}^{\alpha_1\alpha_2}g_{mn},\label{eta-inj-o}\\
\kappa^{\alpha\alpha_1 \alpha_2}_{\text{Inj}}=& - {\pi\nu_c\hbar}\int [d \bm{k} ]\sum_{m \neq n}\frac{\Omega^2}{(\hbar\Omega -\varepsilon_{mn})^2 +\varsigma^2} \Delta_{mn}^{\alpha} \Omega_{mn}^{\alpha_1\alpha_2} g_{mn},\label{kappa-inj-o}\\
\eta^{\alpha\alpha_1 \alpha_1}_{\text{Sh}}=& \varsigma\pi \nu_c\int [d \bm{k} ]\sum_{m\neq n}\Omega^2 \delta(\hbar\Omega -\varepsilon_{mn})R_{mn}^{\alpha\alpha_1} G_{mn}^{\alpha_1\alpha_1} g_{mn},\label{eta-shift-o}\\
\kappa^{\alpha\alpha_1 \alpha_2}_{\text{Sh}}=&-\frac{\pi\nu_c}{2}\int [d \bm{k} ]\sum_{m\neq n} \Omega^2\delta(\hbar\Omega -\varepsilon_{mn})(R_{mn}^{\alpha, +}\lvert\mathcal{A}_{nm}^{+}\rvert^2 -R_{mn}^{\alpha, -}\lvert\mathcal{A}_{nm}^{-}\rvert^2) g_{mn},\label{kappa-shift-o}\\
\eta^{\alpha\alpha_1 \alpha_1}_{\text{Rec}}=&\varsigma 2 \nu_c\int [d \bm{k} ]\sum_{m\neq n}\frac{\Omega^2 }{\hbar\Omega - \varepsilon_{mn}}G_{mn}^{\alpha_1\alpha_2}\partial^{\alpha}g_{mn},\label{eta-rect}\\
\kappa^{\alpha\alpha_1 \alpha_1}_{\text{Rec}}=&-\frac{\nu_c}{2}\int [d \bm{k} ]\sum_{m\neq n} \frac{\Omega^2}{\hbar\Omega -\varepsilon_{mn}}\Omega_{mn}^{\alpha_1\alpha_2}\partial^{\alpha}g_{mn}.\label{kappa-rect}
\end{align}
In which  $\varsigma = 1$ for $\alpha_1 = \alpha_2$ and $\varsigma = -1$ for $\alpha_1 \neq \alpha_2$.

\subsection{Some derivations for the symmetry analysis}

According to \Eq{berry-cone}, the element of the Berry connection matrix is
\begin{equation}
\mathcal{A}_{\bm{k}mn}^{\alpha} = i\sum_{p}(U^{\dagger}_{\bm{k}})_{mp} \frac{\partial (U_{\bm{k}})_{pn}}{\partial k_{\alpha}}=i\sum_{p}U_{\bm{k}pm}^{*} \frac{\partial U_{\bm{k}pn}}{\partial k_{\alpha}}.
\label{Akmn-supp}
\end{equation}
The $\mathcal{T}^\prime$ symmetry gives a constraint on the Berry connection
\begin{equation}
\mathcal{A}_{\bm{k}mn}^{\alpha} = i\sum_{p}(U^{*}_{-\bm{k}})_{mp} \frac{\partial (U_{-\bm{k}}^{T})_{pn}}{\partial k_{\alpha}}
= i\sum_{p}U_{-\bm{k}pm}^{*} \frac{\partial U_{-\bm{k}np}}{\partial k_{\alpha}}=\mathcal{A}_{-\bm{k}nm}^{\alpha}.
\label{bctrans-1}
\end{equation}
By use of \Eq{bctrans-1}, the Berry curvature satisfies the relation (see Eq. (\ref{Omegamtau})):
\begin{eqnarray}
\Omega_{\bm{k}m}^{\tau} &=& \frac{i}{2}\epsilon^{\alpha\gamma\tau}\sum_{n\neq m}(\mathcal{A}_{\bm{k}mn}^{\alpha} \mathcal{A}_{\bm{k}  nm}^{\gamma}-\mathcal{A}_{\bm{k}mn}^{\gamma} \mathcal{A}_{\bm{k}  nm}^{\alpha}) \notag\\
 &=& \frac{i}{2}\epsilon^{\alpha\gamma\tau}\sum_{n\neq m}(\mathcal{A}_{-\bm{k}nm}^{\alpha} \mathcal{A}_{-\bm{k}  mn}^{\gamma}-\mathcal{A}_{-\bm{k}nm}^{\gamma} \mathcal{A}_{-\bm{k}  mn}^{\alpha}) \notag\\
 &=&-\Omega_{-\bm{k}m}^{\tau}.
\label{bcurv-trans-supp}
\end{eqnarray}
For the point-group symmetry transformations $\mathcal{M}$, its eigenvalues $\varepsilon_{n\bm{k}} = \varepsilon_{n\mathcal{M}^{-1}\bm{k}}$, the Berry connection transforms as
\begin{eqnarray}
\mathcal{A}_{\bm{k}mn}^{\alpha} &=& i\sum_{p}[(\mathcal{M}U)^{\dagger}_{\mathcal{M}^{-1}\bm{k}}]_{mp} \frac{\partial (\mathcal{M}U_{\mathcal{M}^{-1}\bm{k}})_{pn}}{\partial k_{\alpha}}  \notag\\
&=& i\sum_{p}\frac{(\partial M^{-1}\bm{k})^\beta}{\partial \bm{k}^{\alpha}}[(\mathcal{M}U)^{\dagger}_{\mathcal{M}^{-1}\bm{k}}]_{mp} \frac{\partial (\mathcal{M}U_{\mathcal{M}^{-1}\bm{k}})_{pn}}{\partial (\mathcal{M}^{-1} \bm{k})_{\beta}}  \notag\\
&=&\mathcal{M}_{\alpha\beta}\mathcal{A}_{\mathcal{M}^{-1}\bm{k}mn}^{\beta}.
\label{Akmn-supp-1}
\end{eqnarray}
The derivative operation transforms as
\begin{equation}
\frac{\partial g_{\mathcal{M}^{-1}\bm{k}m}}{\partial \bm{k}_{\alpha}}=\frac{(\partial M^{-1}\bm{k})^\beta}{\partial \bm{k}^{\alpha}}\frac{\partial g_{\mathcal{M}^{-1}\bm{k}m}}{\partial \bm{k}_{\alpha}}=\mathcal{M}_{\alpha\beta}\frac{\partial g_{\mathcal{M}^{-1}\bm{k}m}}{\partial \bm{k}_{\alpha}}.
\label{partialM-supp}
\end{equation}
\end{widetext}

%

\end{document}